\documentclass[12pt]{article}

\usepackage{a4wide}
\usepackage{amsmath}
\usepackage{graphics}
\usepackage{graphicx}
\usepackage{epsfig}
\usepackage{dcolumn}
\usepackage{bm}
\usepackage{rotating}
\newcommand{\D}{d}

\newcommand{\aaa}{\mbox{\boldmath$a$}}

\newcommand{\AAA}{\mbox{\boldmath$A$}}
\newcommand{\bb}{\mbox{\boldmath$b$}}
\newcommand{\BB}{\mbox{\boldmath$B$}}
\newcommand{\cc}{\mbox{\boldmath$c$}}

\newcommand{\vel}{\mbox{\boldmath$v$}} 
\newcommand{\CC}{\mbox{\boldmath$C$}}

\newcommand{\ee}{\mbox{\boldmath$e$}}

\newcommand{\ff}{\mbox{\boldmath$f$}}  

\newcommand{\gggg}{\mbox{\boldmath$g$}}

\newcommand{\kk}{\mbox{\boldmath$k$}}
\newcommand{\KK}{\mbox{\boldmath$K$}}
\newcommand{\LL}{\mbox{\boldmath$L$}}
\newcommand{\mm}{\mbox{\boldmath$m$}}
\newcommand{\MM}{\mbox{\boldmath$M$}}
\newcommand{\nn}{\mbox{\boldmath$n$}}

\newcommand{\qq}{\mbox{\boldmath$q$}}

\newcommand{\uu}{\mbox{\boldmath$u$}}

\newcommand{\vv}{\mbox{\boldmath$v$}}
\newcommand{\ww}{\mbox{\boldmath$w$}}

\newcommand{\xx}{\mbox{\boldmath$x$}}

\newcommand{\UNIT}{\mbox{\boldmath$1$}}
\newcommand{\bONE}{\mbox{\boldmath$1$}}

\newcommand{\bnabla}{\mbox{\boldmath$\nabla$}}
\newcommand{\bxi}{\mbox{\boldmath$\xi$}}

\newcommand{\bsigma}{\mbox{\boldmath$\sigma$}}
\newcommand{\bkappa}{\mbox{\boldmath$\kappa$}}  

\newcommand{\grad}{\mbox{\boldmath$\nabla$}}

\newcommand{\bDelta}{\mbox{\boldmath$\Delta$}}

\begin{document}

\title{Invariance correction to Grad's  equations:\\
Where to go beyond approximations?}
\author{Alexander N. Gorban\\
    Department of Mathematics, University of Leicester, LE1 7RH
    Leicester,  UK
    \vspace*{1cm}\\
Iliya V. Karlin\thanks{Corresponding author}\\
ETH Z\"{u}rich,
    Institute of Energy Technology,
    CH-8092 Z\"{u}rich,
    Switzerland}

  \date{\today}

  \maketitle

\begin{abstract}

We review some recent developments of Grad's approach to solving the
Boltzmann equation and creating reduced description. The method of
invariant manifold is put forward as a unified principle to
establish corrections to Grad's equations. A consistent derivation
of regularized Grad's equations in the framework the method of
invariant manifold is given. A new class of kinetic models to lift
the finite-moment description to a kinetic theory in the whole space
is established. Relations of Grad's approach to modern mesoscopic
integrators such as the entropic lattice Boltzmann method are also
discussed.

\end{abstract}


\section{Introduction}
\label{intro} There has been a long-standing quest for improving on
the Grad $13$-moment approximation \cite{Grad49}. In particular,
such an improvement is needed to study the interplay between
hydrodynamics and kinetics in the domain of moderate Knudsen
numbers, in particular, simulations of flows at a micrometer scale
in so-called micro-electro-mechanical systems (MEMS)
\cite{Karniadakis01}. The recent renewed interest to this topic is
consistent with the current trend in computational fluid mechanics
to use minimal kinetic models instead of more traditional numerical
schemes for hydrodynamic equations.

Let us recall the famous Grad's $10$-moment and $13$-moment
approximations for the distribution function:

\begin{equation}
\label{10} f=f^{(0)} \left\{
1+\frac{1}{p}\bsigma:\left(\cc\cc-\frac{1}{3}\UNIT
c^2\right)\right\},
\end{equation}

\begin{equation}
f=f^{(0)} \left\{1+
\frac{1}{p}\bsigma:\left(\cc\cc-\frac{1}{3}\UNIT c^2\right)
 + \frac{4}{5pv_T}\qq\cdot\cc\left(c^2-\frac{5}{2}\right)
\right\}. \label{13}
\end{equation}
Here, as usual, $f^{(0)}$ is the local Maxwellian,
$\cc=v_T^{-1}(\vv-\uu)$ is the ``peculiar'' velocity, $\uu$ is the
local flow velocity,  $v_T=\sqrt{m/2k_{\rm B}T}$ is thermal
velocity, $p$ is scalar pressure,  $\bsigma$ is the nonequilibrium
stress tensor, and $\qq$ is the heat flux,

\begin{eqnarray*}
\sigma _{ik}(f)&=&\int  \left[m(v_{i}-u_{i})(v_{k}-u_{k})-{1\over
3}\delta _{ik}m\left({\vv}-{\uu}\right)^{2}\right]fd\vv, \\
q_i(f)&=&\int  \left[{m\over 2}(v_{i}-u_{i}) \left(
({\vv}-{\uu})^{2}-{5k_{B}T\over m} \right)\right]fd\vv.
\end{eqnarray*}

Technically, in Grad's original approach, parametric families
(\ref{10}) and (\ref{13}) were introduced as truncated Hermite
polynomial expansions of the distribution function around local
Maxwellians. However, it is much more attractive to view Grad's
distributions as parametrically specified sub-manifolds
(``surfaces") in the larger space of distribution functions.
Grad's  method has given start to a host of new methods focused
around the hard question of nonequilibrium statistical mechanics:
How to effectively reduce the microscopic  to a macroscopic
description? This review is devoted to some selected instances of
this question.

\section{Grad's method and beyond}

``Grad's legacy'' (where and how to go beyond the $13$-moment
approximation?)  was interpreted and extended in different ways by
many authors.  Let us mention those which are most relevant to the
present discussion.

\subsection*{Quasi-equilibrium approximation}

\textit{Quasi-equilibrium approximation} (or {\it maximum entropy
  approximation}) in the application to the Boltzmann equation was
established in the sixties by several authors, in particular, by
Kogan \cite{Kogan65} and Lewis \cite{Lewis67}, though we note that
it was already mentioned by Grad himself, and also by Koga (cf.\
\cite{Lewis67}). A detailed discussion geometrical aspects of
quasi-equilibria was given in \cite{obkhod}. The construction is
based on solving the conditional maximization problem: For the
concave functional $S=-k_{\rm B}\int f\ln f d\vv$ (local entropy
density) and for given distinguished linear functionals $M(f)$,
find

\begin{equation} \label{smax} S\to\max,\ M(f)=M.  \end{equation}

The solution in terms of Lagrange multipliers (dual variables)
$\Lambda$ is written as

\begin{equation}\label{dual}
f=\exp\left\{\sum_k\Lambda_kD_fM_k\right\}.
\end{equation}

If $M_k(f)=\int m_k(\vv)fd\vv$, then we have

\begin{equation}\label{dual2} f=\exp\left\{\sum_k\Lambda_km_k\right\}.
\end{equation}

If $M=M_0=\int\{1,\vv,v^2\}fd\vv$, the parametric set (\ref{dual})
coincides with the set of local Maxwellians.  If
$M=\int\{1,\vv,\vv\vv\}fd\vv$ this is the $10$-moment
quasi-equilibrium approximation, whose expansion to linear order
in $\bsigma/p$ coincides with Grad's $10$-moment approximation
(\ref{10}).  Though almost pedantic, some attention is required
when proceeding to the $13$-moment case: Functions (\ref{dual})
are defined in terms of dual variables.  What is not always well
defined is the moment chart (or moment parameterization) of these
sets, or, $\Lambda(M)$, and a regularization of divergent
integrals is required. This can be done either by restricting the
velocity integration domain to a large ball $v\le\sqrt{E}$, where
$E$ is the total kinetic energy of the gas in a container
\cite{Kogan65}, or by introducing a higher-order even velocity
polynomial  (at a price of an extra variable).  In the first case
of regularization it is possible to use the smallness of
$\qq/pv_T$ to expand the regularized distribution and send the
radius of the ball to infinity to end up with the $13$-moment
Grad's approximation (\ref{13}).

A particularly useful version of entropic methods was introduced in
\cite{Karlin86} for chemically reacting gas mixtures, and
discussed in some detail for a single-component gas in
\cite{GK91,GK96} ({\it triangle entropy method}).  It can be viewed
either as a stepwise realization of the basic maximization problem, or
(better) as a self-consistent recipe.  Let us split the totality of
distinguished macroscopic variables $M$ into $M'$ and $M''$, where
$M'$ are linear functionals for which we can solve explicitly the
problem (\ref{smax}) (``easy variables''), and where $M''$ are
``difficult variables''. The difficult variables may be even
nonlinear in the distribution function, for
example scattering rates (see below), so that even the statement
of the problem of the entropy maximization can cause difficulties.
The triangle entropy method allows to
construct quasi-equilibria even in these cases.
 Let us denote $f(M')$ the quasi-equilibrium
corresponding to the easy variables.  Then the triangle
quasi-equilibrium for the difficult variables is found as follows:
expand the entropy functional up to quadratic order around $f(M')$,
and find maximum under the conditions that (i) Easy variables $M'$ are
not changed, and (ii) Difficult variables are fixed to linear order.
That is,

\begin{eqnarray}
 \Delta S(\delta f)&=&S(f(M'))-k_{\rm B}\int [\ln f(M')+1]
\delta f d\vv+ \frac{1}{2}k_{\rm B} \int f(M')^{-1}\delta
f^2d\vv\to\max,\nonumber \\ &{\rm (i)}& M'(\delta f)=0,\nonumber
\\&{\rm (ii)}& D_f M''\big|_{f(M')}(\delta f)=\delta
M''.\label{trianglesmax}
\end{eqnarray}
Maximization here is with respect to $\delta f$, nonlinear
parametric dependence on $M'$ is not varied.  The nice property is
that (\ref{trianglesmax}) is always solvable in closed form, the
resulting triangle quasi-equilibrium,

\begin{equation}
 \label{triangle}
f(M',\delta M'')=f(M')+\delta f(M', \delta M''),
\end{equation}
depends linearly on $\delta M''$ and nonlinearly on $M'$ (so the
overall dependence is \textit{quasi-linear}).  If $M'$ are the five
hydrodynamic fields, and if $M''$ are $\int \vv\vv f d\vv$ or $\int
\{\vv\vv,\vv v^2\}fd\vv$ (both easy and difficult variables are linear
in this example), then (\ref{triangle}) are Grad's $10$- and
$13$-moment distributions, respectively \cite{GK91}.

The advantage of the quasi-equilibrium approximations is that they
are equipped naturally with the \textit{thermodynamic
parameterization} \cite{GK92}.  The structure of the thermodynamic
parameterization assumes specification of the projector $P$ onto
the tangent bundle of the quasi-equilibrium manifold.  For
quasi-equilibria, this is (we stick to the case (\ref{dual2}) for
simplicity):

\begin{equation}
\label{proj} PJ=\sum_k\frac{\partial f(M)}{\partial M_k}\int m_k
Jd\vv.
\end{equation}

The purpose of projector $P$ is to define dynamics along the manifold.
Namely, if we write the Boltzmann equation,

\begin{equation}
\label{BE}
D_tf=J(f)=-(\vv-\uu)\cdot\grad f+Q(f,f), \end{equation}
where $Q(f,f)$ is the Boltzmann collision integral, and
$D_t=\partial_t+\uu\cdot\grad$ is the time derivative in the co-moving
reference system, then the vector field attached to each state on the
quasi-equilibrium manifold (or the \textit{microscopic time
  derivative}) is:

\begin{equation}
\label{micro} D_t^{\rm micro}f(M)=J(f(M)).
\end{equation}
Here, we simply evaluate the action of the operator in the right hand
side of the Boltzmann equation (\ref{BE}) on the quasi-equilibrium
distributions.  On the other hand, under the action of the projector
$P$ (\ref{proj}), vectors $J(f(M))$ yield the vector field on the
tangent bundle of the quasi-equilibrium manifold, or the
\textit{macroscopic time derivative}:

\begin{equation}
\label{macro}
D_t^{\rm macro}f(M)=PJ(f(M)).
\end{equation}
The latter can be viewed as a short-hand writing of Grad's equations,
which follow from (\ref{macro}) upon multiplication with $m_k$ and
integration:

\begin{equation}\label{Gradeq}
\partial_t M_k+\uu\cdot\grad M_k=\int m_k PJ(f(M))d\vv.
\end{equation}

One can ask, what is the use of the microscopic time derivative
(\ref{micro}) when only its projected piece, $PJ(f(M))$, contributes
finally to Grad's moment equations (\ref{Gradeq})?  The answer is that
the comparison of the vectors $J(f(M))$ and $PJ(f(M))$ measures how
good the closure (\ref{Gradeq}) really is.  The difference between
$J(f(M))$ and $PJ(f(M))$ is of such a great importance that it
deserves a specific name.  The \textit{defect of invariance} (of the
quasi-equilibrium approximation) is,

\begin{equation}
\label{defect}
\Delta(M)=J(f(M))-PJ(f(M)).
\end{equation}

A moment representation of the defect is also useful: If
$m_1,\dots,m_n$ are the distinguished moments, and $m_{n+1}, \dots$
are the higher-order moments, then the velocity-dependent function is
equivalent to the infinite sequence $\Delta_i(M)$ by taking moments of
(\ref{defect}):

\begin{equation}
\label{defmom}
\Delta_i(M)=\left\{\begin{array}{ll} 0, & i=1,\dots, n\\
\int m_i(J(f(M))-PJ(f(M)))d\vv, & i=n+1, \dots\ \end{array} \right.
\end{equation}
 Levermore \cite{Levermore96} proved
hyperbolicity of maximum entropy approximations. Dual
parameterization of quasi-equilibrium manifolds prove to be an
advantage in numerical realizations (so-called Legendre
integrators, see e.\ g.\ \cite{Ilg2002,Ilg2003,GGK04} in the
context of polymer dynamics).

The quasi-equilibrium approximations reveal most clearly the {\it
time hierarchy assumption} behind the Grad's approach
\cite{Kogan65}: In the fast relaxation, the entropy grows
according to Boltzmann equation until maximum of entropy is
reached on the quasi-equilibrium states. After that the slow
evolution takes place along the manifold of the quasi-equilibrium
states.  {\it Putting this assumption on trial and,
  if needed, improving on it} is the key in seeking corrections to
Grad's approximation.  The trial is the deviations away from zero of
the defect of invariance (\ref{defect}).

In this section, we reviewed the basic structure of Grad's theory, and
indicated that a way beyond a given moment approximation should take
into account the defect of its invariance.  Before proceeding along
this line, let us discuss two other routes, which can be indicated as
``increase the number of variables'' and ``take other variables''.

\subsection*{Many moments approximations}

With a given moment approximation at hand, and without asking the
question, ``How good is this approximation?''  there is only one
option to try to improve on it - to extend the list of
distinguished variables, and to construct another approximation.
This viewpoint dominated earlier studies on moment approximations,
and was followed by many authors, in particular, by M\"uller and
Ruggeri \cite{Muller} and their collaborators, mostly in the
quasi-linear form, using orthogonal functions developments, and
for the Bhatnagar-Gross-Krook model collision integral with
phenomenological dependence of the relaxation parameter.
Convergence to Boltzmann equation is a difficult question; in
fact,  it is not expected that Grad's distributions converge to
solutions of the Boltzmann equation, even pointwise
\cite{Bobylev82}.  On the other hand, as numerical results show
\cite{Muller}, the weak convergence (convergence of the moments)
can be expected in the linear case.  However, without at least
evaluating the defect, Eqs.\ (\ref{defect}) or(\ref{defmom}), the
uncontrollability of approximations with any number of moments
remains.

\subsection*{Scattering rates as independent variables}
Remarkably, Grad's approximations with moments as slow
variables (\ref{10}) and (\ref{13}) (or any other with more moments)
do not contain any molecular information.  This shortcoming is
inherited from the use of simple sets of orthogonal functions in the
original Grad method.  However, if one thinks of using less variables
to capture more physics, then other (non-moment) quantities can be
tried.  In particular, for the pure gas, interesting variables are
{\it scattering rates of moments}, for example,

\begin{equation}
\label{collstress}
\bsigma^{\rm coll}=\int m\vv\vv Q(f,f)d\vv,
\end{equation}

is the scattering rate of the stress tensor, $Q$ is the Boltzmann
collision integral.  The variable $ \bsigma^{\rm coll}$, unlike $
\bsigma$, contains information about the molecular interaction.
Using the triangle entropy method with the five hydrodynamic
variables as easy, and $ \bsigma^{\rm coll}$ as difficult (this is
indeed the case because $ \bsigma^{\rm coll}$ is nonlinear in
$f$), we can construct a ``scattering'' counterpart of the
$10$-moment approximation (\ref{10}):

\begin{equation} \label{10coll} f=f^{(0)} \left\{ 1+\frac{1}{p\mu_0(T)}
\bsigma^{\rm coll}:\left(\cc\cc-\frac{1}{3}\UNIT
c^2\right)R(c^2)\right\},
\end{equation}
where $\mu_0(T)$ is the first Sonine polynomial approximation to
the viscosity coefficient, and the difference from Grad's
$10$-moment approximation is in the dimensionless function $R$.
The function $R$ depends on the particle's interaction, and $R$ is
constant only for Maxwell molecules in which case the present
approximation is equivalent to Grad's approximation (up to
renaming the variables). For hard spheres \cite{GK96},

\begin{equation}
\label{RS}
R= \frac {5\sqrt 2}{16}
\int_0^1 e^{-c^2 t^2}(1-t^4 )\left( c^2 (1-t^2 ) + 2 \right) dt.
\end{equation}

The case when the scattering rate of the heat flux is included (the
counterpart of the $13$-moment approximation), and also a {\it mixed}
version (moments and scattering rates both as distinguished
variables), were studied in Ref.\ \cite{GK91}.

Eventually, the triangle entropy method makes it possible to
handle ``smart'' variables which may be more appropriate to the
physics of the problem at hand rather than plain moments.  The
latter assumes a certain degree of a physical intuition;
furthermore, the uncontrollability of the resulting quasi-linear
quasi-equilibria remains an issue.

\subsection*{Method of invariant manifold}

The general method to derive dynamic corrections on top of
successful initial approximations like Grad's was developed by the
authors \cite{GK92,Kdiss,GK94inv,GKbook1,GKZ2004,GKbook2}. The
essence of the \textit{method of invariant manifold} (MIM) is (i)
to write the invariance condition \textit{in the differential
form} (the microscopic time derivative on the manifold equals the
macroscopic time derivative), and (ii) to solve this equation by
iterations. The choice of the initial approximation is an
important problem. Often, it is convenient to start from the
quasi-equilibrium manifold (this will be our choice below in this
paper). However, the choice of the initial manifold in MIM is not
restricted to quasi-equilibria. The typical example gives us the
famous Tamm--Mott-Smith approximation for the strong shock wave
(see discussion in \cite{GK92,GKbook2,GK04b}). Strictly speaking,
the method of invariant manifold can be applied in order to refine
any initial approximation compliant with some transversality
conditions.

\subsubsection*{Correction to local Maxwell manifold}

In order to explain the two steps of the MIM, we shall consider first,
for the purpose of illustration, the invariance correction to the
local Maxwell approximation to all orders in Knudsen number.

The main idea is to pose the problem of a finding a correction to
Euler's hydrodynamics in such a way that Knudsen number expansions do
not appear as the necessary element of the analysis.  This will be
possible by using Newton method instead of Taylor expansions to get
such correction.

The starting point is the manifold of local Maxwell distribution
functions (LM) $f_0(n,\mbox{\boldmath$u$},T;\vv)$, where $\vv$ is
particle's velocity, and $n$, $\uu$, and $T$ are local number density,
average velocity, and temperature.  As appropriate to our approach, we
first check the invariance of the LM manifold.  Projector (\ref{proj})
on the LM manifold is:

\begin{equation}
\label{proj0} P_0J=\frac{f_0}{n}\left[\int Jd\cc+2\cc\cdot \int
\cc J d\cc +\frac{2}{3}\left(c^2-\frac{3}{2}\right)\int
\left(c^2-\frac{3}{2}\right) J
 d\cc \right],
\end{equation}

Computing the microscopic time derivative (\ref{micro}) on the LM
states, projecting it with $P_0$ (\ref{proj0}) to get the
macroscopic time derivative (the time derivative due to Euler's
equations) (\ref{macro}), and subtracting the second out of the
first, we evaluate the invariance defect of the LM manifold
(\ref{defect}):

\begin{equation}
\label{delta0}\Delta(f_0)=J(f_0)-P_0 J(f_0 )=-f_0
\left[2\nabla\uu:\left(\cc\cc -\frac{1}{3}\mbox{I}c^2 \right)+v_T
\frac{\nabla T}{T}\cdot\cc\left(c^2 -\frac{5}{2}\right)\right].
\end{equation}
The defect is not equal to zero as long as there average velocity and
the temperature vary in space, as expected.  Note that the defect is
neither small or large by itself.

To find the correction to the LM manifold, we write the invariance
condition,

\begin{equation}
\label{inv} \Delta(f)=J(f)-PJ(f)=0,
\end{equation}
and consider it as an \textit{equation} to be solved with the
initial approximation $f_0$ for the manifold and $P_0$ for the
projector since \textit{both} are unknown in the (\ref{inv}). This
might seem too much to require, however, the well-posedness of the
problem is restored once the \textit{additional requirement} that
the manifold we are looking for should be the manifold of slow
motions is invoked (see Ref.\ \cite{GK94inv} for details). Here we
will consider the first iteration.

Upon substitution of $P_0$ in place of $P$, and of $f_1 = f_0 + \delta f$
in place of $f$ in equation (\ref{inv}), and after the
linearization in $\delta f$, we get

\begin{equation}
\label{it}
L_{f_0}\delta f +(P_0 -1)(\vv-\uu)\cdot \nabla \delta f + \Delta (f_0 ) = 0,
\end{equation}
where $L_{f_0}$ is the linearized collision integral (linearization in
the local Maxwell state, and we keep indicating the linearization
point for reasons to be seen later).  Equation (\ref{it}) has to be
solved subject to the condition,

\begin{equation}
\label{cond} P_0 \delta f = 0.
\end{equation}

The linear \textit{equation of the first iteration} (\ref{it}) is the
most important object in our theory.  Indeed, it does not contain at
all the smallness parameter, and, in fact, was obtained without
assumption of a small Knudsen number.  If, however, the Knudsen number
is introduced into equation (\ref{it}) by the usual rescaling,

\begin{equation}
\label{kn} L_{f_0}\to \frac{1}{\epsilon}L_{f_0},
\end{equation}
then the first-order in $\epsilon$ solution, $\delta f_1\simeq\epsilon
\delta f_1^{(1)}$ is found from the integral equation,

\begin{equation}
\label{CE}L_{f_0} \delta f_1^{(1)}= -\Delta (f_0 ),
\end{equation}
which has to be solved subject to the condition $P_0\delta f_1^{(1)}$.
It is readily checked that equation (\ref{CE}) is just the equation of
the first approximation of the Chapman-Enskog method, which is thus
recovered as the special case of MIM in the collision-dominated limit.

The invariance equation for the first correction (\ref{it}) is
more complicated than the Chapman-Enskog equation (\ref{CE})
because it also contains the spatial derivatives (though it is
much \textit{less} complicated than the linearized Boltzmann
equation because the there is no time dependence in the equation
(\ref{it})).  Methods to treat equations of the type (\ref{inv})
has been also developed in \cite{Kdiss,GK94inv,GKbook2} and worked
out for many kinetic systems (not only for the Boltzmann
equation). Here we consider, for the purpose of illustration, the
small deviations around the global equilibrium in $3D$.  We shall
treat equation (\ref{it}) in such a way that the Knudsen number
will appear explicitly only at the latest stages of the
computations.

We denote as $F_0$ the global equilibrium with the equilibrium values
of the hydrodynamic quantities, $n_0 $, $\uu_0 =0 $, and $T_0$.
Deviations are $\delta n $, $\delta \uu$, and $\delta T$.  We also
introduce reduced deviations, \[\Delta n = \delta n/n_0,\ \Delta
\uu=\delta \uu/v_T^0, \Delta T = \delta T/T_0,\] where $v_T^0 $ is
thermal velocity in the equilibrium.

We seek the invariance correction,
\begin{equation}\label{corrlin}
  f_1 = F_0 (1+\varphi_0 + \varphi_1 ),
\end{equation}
where
\begin{equation}\label{maxwellin}
\varphi_0 = \Delta n + 2\Delta\uu\cdot\cc
\unboldmath +\Delta T \left(c^2 - \frac{3}{2}\right).
\end{equation}
comes from the linearization of the local Maxwellian around $F_0$, and
where $\varphi_1 $ is the unknown function to be found from equation
invariance equation (\ref{it}).  In order to find $\varphi_1 $, we
apply a Galerkin approximation in order to achieve a
finite-dimensional approximation of the linear collision operator
\cite{GK94inv}, which amounts to setting

\begin{equation} \label{ansatz} \varphi_1 = \AAA(\xx)\cdot
\cc\left(c^2 - \frac{5}{2}\right)
+\BB(\xx):\left(\cc\cc- \frac{1}{3}Ic^2 \right).  \end{equation}

Our goal is to derive functions $\AAA$ and $\BB$ from a linearized
version of equation (\ref{it}).  Knowing $\AAA$ and $\BB$, we get
the following expressions for shear stress tensor $\bsigma$ and
heat flux vector $\qq$:

\begin{equation} \label{sigma}
\bsigma = p_0 \BB,\ \qq=\frac{5}{4}p_0 v_T^0 \AAA,
\end{equation}
where $p_0 $ is the equilibrium pressure.

Linearizing equation (\ref{it}) in $F_0 $, substituting $\varphi_1$
(\ref{ansatz}), and switching to the Fourier-transformed in space
variables, we derive the set of linear algebraic equations for the
Fourier image of the functions $\AAA$ and $\BB$ (which we denote as
$\aaa_k$, and $\bb_k$, respectively):

\begin{eqnarray}
\label{3d} \frac{5p_0}{3\mu_0}\aaa_k+iv_T^0 \bb_k\cdot\kk &
=&-\frac{5}{2}iv_T^0\kk\tau_k;\\\nonumber
\frac{p_0}{\mu_0}\bb_k+iv_T^0\overline{\kk\aaa_k }
&=&-2iv_T^0\overline{\kk\mbox{\boldmath$\gamma $}_k},
\end{eqnarray}
where $i=\sqrt{-1}$, $\kk$ is a wave vector, $\mu_0$ is the first
Sonine polynomial approximation of the shear viscosity
coefficient, $\tau_k$ and \boldmath$\gamma$\unboldmath$_k$ are
Fourier images $\Delta T$, and $\Delta\uu$, respectively, and the
over-bar denotes a symmetric traceless dyad.

Introducing dimensionless the reduced wave vector,
\[\bkappa=\frac{v_T^0\mu_0}{p_0}\kk,\]
solution to equation (\ref{3d}) may be written:
\begin{eqnarray}
\bb_k&=&-\frac{10i\overline{\mbox{\boldmath$\gamma $}_k
\bkappa}}{3[(5/3)+(1/2)\kappa^2]}+ \frac{5i(\mbox{\boldmath$\gamma
$}_k
\cdot\bkappa)\overline{\bkappa\bkappa}}{3[(5/3)+(1/2)f^2][5+2\kappa^2]}-
\frac{15\tau_k\overline{\bkappa\bkappa}}{2[5+2\kappa^2]},\nonumber\\
\aaa_k&=&-\frac{15i\bkappa\tau_k}{2[5+2\kappa^2]}-
\frac{5[\bkappa(\mbox{\boldmath$\gamma $}_k
\cdot\bkappa)+\mbox{\boldmath$\gamma
$}_kf^2(5+2\kappa^2)]}{3[5+2\kappa^2][(5/3)+(1/2)\kappa^2]}.
\label{SOLUTION3D}
\end{eqnarray}
With the Fourier-image of the fluxes (\ref{sigma}),
\[\bsigma_k = p_0
\bb_k,\ \qq=\frac{5}{4}p_0 v_T^0 \aaa_k, \]

which have to be used to close the Fourier-transformed linear balance
equations, functions (\ref{SOLUTION3D}) concludes our computation of
the dynamic correction to the linearized local Maxwellian.  Note that
due to the non-polynomial in $\kappa$ contributions, the resulting
linear hydrodynamics is highly nonlocal.  This is, of course, not
surprising because no small Knudsen number expansions truncated to
some order ever appeared.

Let us briefly consider the new hydrodynamic equations specializing to
the one-dimensional case.  Taking the $z$-axis for the propagation
direction, and denoting $k_z $ as $k$, $\gamma $ as $\gamma_z $, we
obtain in (\ref{SOLUTION3D}) the Fourier images of $a = a_z $ and $b =
b_{zz}$ (full notation are restored here):

\begin{eqnarray}
\label{1d-dim} a_k & = & - \frac {\frac{3}{2} p_0^{-1}\mu_0 v_T^0
ik \tau_k +\frac{4}{5}  p_0^{-2}\mu_0^2 (v_T^0 )^2 k^2  \gamma_k
}{1+\frac{2}{5} p_0^{-2}\mu_0^2 (v_T^0 )^2 k^2 },\\\nonumber b_k &
= & -\frac {\frac{4}{3} p_0^{-1}\mu_0 v_T^0 ik \gamma_k +
p_0^{-2}\mu_0^2 (v_T^0 )^2 k^2 \tau_k } {1+\frac{2}{5}
p_0^{-2}\mu_0^2 (v_T^0 )^2 k^2 }.
\end{eqnarray}
These expressions close the linearized balance equations,

\begin{eqnarray}
\label{hydro} \frac {1}{v_T^0 }\partial_t \nu_k +  ik \gamma_k & =
& 0, \\\nonumber \frac {2}{v_T^0 }  \partial_t \gamma_k +
ik(\tau_k +\nu_k ) + ik b_k & = & 0, \\\nonumber \frac {3}{2v_T^0
}\partial_t \tau +  ik\gamma_k + \frac{5}{4} ika_k & = & 0.
\end{eqnarray}

In order to restore the Knudsen number in (\ref{1d-dim}), we
introduce $l_{\textrm{m.f.p.}} = v_T^0 \mu_0 /p_0 $ (the quantity
$l_{\textrm{m.f.p.}}$ is of the order of the mean free path), and
we also introduce a hydrodynamic scale $l_{h}$, so that $k =
\kappa /l_h $, where $\kappa$ is dimensionless.  With this, we
obtain in the equation (\ref{1d-dim}):

\begin{eqnarray}
\label{1d} a_{\kappa}  &=& - \frac {\frac{3}{2} i\epsilon\kappa
\tau_{\kappa} + \frac{4}{5}\epsilon^{2} \kappa^ {2}
\gamma_{\kappa} } {1 + \frac{2}{5}\epsilon^{2} \kappa^{2}
},\\\nonumber b_{\kappa}&=&- \frac {\frac{4}{3}i\epsilon\kappa
\gamma_{\kappa} + \epsilon^2 \kappa^2 \tau_{\kappa} } {1 +
\frac{2}{5}\epsilon^2 \kappa^2   },
\end{eqnarray}
where $\epsilon=l_c /l_h $ is the Knudsen number. In the limit
$\epsilon\rightarrow 0 $, equation (\ref{1d})reduces to the
familiar Navier-Stokes-Fourier expressions:
\[\sigma_{zz}=-\frac{4}{3} \mu_0
\partial_z \delta u_z,\ q_z = -\lambda_0 \partial_z \delta T\] where
$\lambda_0 = 15k_B \mu_0 /4m $ is the first Sonine polynomial
approximation of the thermal conductivity.

In order to find out a result of non-polynomial behavior
(\ref{1d}), it is most informative to calculate a dispersion
relation for planar waves.  It is worthwhile introducing
dimensionless frequency $\lambda = \omega l_h / v_T^0 $, where
$\omega$ is a complex variable of a wave $\sim \exp(\omega t +
ikz)$ ($\mbox{Re}\omega $ is the damping rate, and
$\mbox{Im}\omega $ is the circular frequency).  Making use of
(\ref{hydro}) and (\ref{1d}), writing $\epsilon=1$, we obtain the
following dispersion relation $\lambda(\kappa)$:

\begin{equation}
\label{disp} 12(1+\frac{2}{5}\kappa^2 )^2 \lambda^3 + 23\kappa^2
(1+\frac{2}{5}\kappa^2 )\lambda^2 +2\kappa^2 (5+5\kappa^2
+\frac{6}{5}\kappa^4 )\lambda + \frac{15}{2}\kappa^4
(1+\frac{2}{5}\kappa^2 )=0.
\end{equation}

\begin{figure}[t]
\centering{
\frame{\includegraphics[height=.30\textheight]{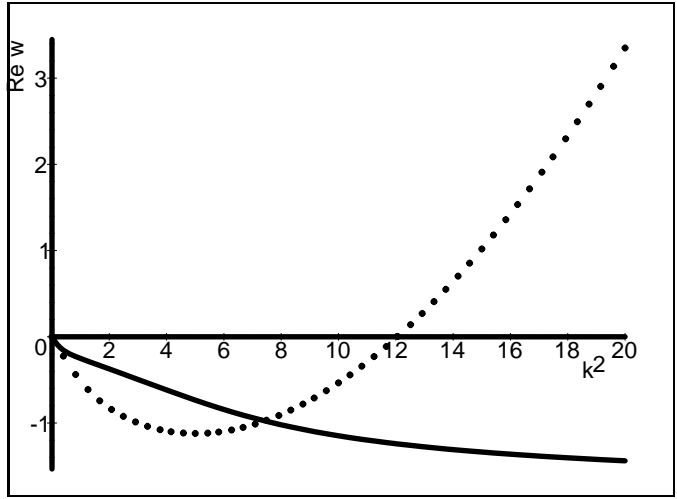}}
\caption{\label{dispersion} Attenuation rate of sound waves. Dotts:
Burnett approximation. Bobylev's instability occurs when the curve
intersects the horizontal axis. Solid: First iteration of the Newton
method on the invariance equation.}}
\end{figure}

Fig.\ \ref{dispersion} presents a dependence ${\rm Re}\lambda
(\kappa^2 )$ for acoustic waves obtained from (\ref{disp}) and for
the Burnett approximation \cite{Bobylev82}.  The violation in the
latter occurs when the curve crosses the horizontal axis.  In
contrast to the Burnett approximation \cite{Bobylev82}, the
acoustic spectrum (\ref{disp}) is stable for all $\kappa $.
Moreover, ${\rm Re}\lambda$ demonstrates a finite limit, as
$\kappa \rightarrow \infty $ (so-called ``Rosenau saturation"
\cite{Slemrod98}).

 The example considered demonstrates how to apply the method of
invariant manifold in the simplest case of the initial manifold.
Let us now switch back to another case of the initial manifold,
the Grad's $13$-moment approximation.

\subsubsection*{Invariance correction to $13$-moment manifold}

As we`ve said before, MIM is able to address the invariance
correction, in principle, to any interesting initial
approximation, so it may be not surprising that the next candidate
after the local Maxwell manifold are manifolds of Grad's
distributions.  The problem of finding the invariance correction
to the moment approximations was first addressed in Ref.\
\cite{Karlin98}.  Without repeating computations of the Ref.\
\cite{Karlin98}, our objective here is to explain why the
correction to the Grad's manifolds is a distinguished case.  Let
us start with the quasi-equilibrium of a generic form, and compute
the first iteration of the invariance equation,

\begin{equation}\label{itqe}
  (1-P)L_{f(M)}\delta f+(P-1)(\vv-\uu)\cdot\grad\delta
  f+\Delta(M)=0,
\end{equation}

where $P$ is the quasi-equilibrium projector (\ref{proj}), $\Delta(M)$
is the invariance defect of the quasi-equilibrium approximation, and
$L_{f(M)}$ is the collision integral, \textit{linearized in the
  quasi-equilibrium}.  Notice that the latter object is not well
studied in the classical theory of the Boltzmann equation, where most
of the reduction problems are using the collision integral, linearized
in the local equilibrium.  A simplification of equation (\ref{itqe})
happens when we look for Grad's quasi-linear approximations, then,
\textit{to the linear order accuracy in the higher-order moments}, we
can replace,

\begin{equation}
\label{replace} L_{f(M)}\to L_{f_0},\end{equation}

in the equation (\ref{itqe}), where $L_{f_0}$ is the usual linearized
collision integral in the local Maxwell state.

Let us proceed further with evaluation of the other terms in the
equation (\ref{itqe}).  The projector, corresponding to the
$13$-moment Grad's approximation, reads,

\begin{equation}
\label{PG}
 P_{13}=P_0+\Pi_{13},
\end{equation}

where $P_0$ is the local-equilibrium projector (\ref{proj0}), and
where $\Pi_{13}$ acts as follows:

\begin{equation}
\Pi_{13}J\!=\!\frac{f_{0}}{n}\!\left\{{\sf Y}:\int{\sf Y}J d{\vv}
+{\bf Z}\cdot\int{\bf Z}Jd{\vv}\right\}, \label{dPG}
\end{equation}

Here ${\sf Y}=\sqrt{2}\overline{{\bf c}{\bf c}}$,
and ${\bf Z}=\frac{2}{\sqrt{5}}{\bf c}\left(c^2 -\frac{5}{2}\right)$, are
peculiar velocity polynomials forming the $13$-moment approximation.

Computing the defect of invariance of the $13$ moment approximation to
the linear order in the $\varphi_{13}$, we see, that there are two
contribution, local (containing the linearized collision integral),
and the nonlocal (containing the free flight operator),

\begin{eqnarray}
\label{D13M} \Delta_{13}&=&\Delta_{13}^{{\rm loc}}+ \Delta^{{\rm
nloc}}_{13},\\\nonumber \Delta_{13}^{{\rm
loc}}&=&(1-\Pi_{13})[L_{f_0}f_0 \varphi_{13}],\\\nonumber
\Delta^{{\rm nloc}}_{13}&=&(1-P_{13}) [-(\vv-\uu)\cdot\nabla
f_0(1+\varphi_{13})].
\end{eqnarray}

Before proceeding any further, we shall discuss the physical
significance of the defect (\ref{D13M}) because it is the first
instance where classical methods, like the Chapman-Enskog method,
become inapplicable.

The nonzero defect of invariance of any manifold reveals the
following: The solution to the Boltzmann equation with the initial
condition on the manifold leaves this manifold at $t>0$.  The two
parts of the defect correspond to two different mechanisms
responsible for this to happen.  The local defect is not equal to
zero whenever the polynomials ${\sf Y}$ and ${\bf Z}$, forming the
Grad manifold, are not eigenvectors of the linearized collision
integral.  This is distinct from the dynamic noninvariance of
local Maxwellians, where in the latter case the local defect is
equal to zero whatever the collision model is chosen.  For the
Grad approximation, $\Delta_{13}^{{\rm
    loc}}=0$ in only two (commonly known) cases, i. e. for Maxwell molecules and for the
Bhatnagar-Gross-Krook model (BGK).

It is important to recognize that, whenever the local defect is
not equal to zero, the initial manifold has to be \textit{first}
corrected locally, in order to bring it closer to the slowest
eigenspace of the collision operator, and \textit{before} any
nonlocal corrections due to $ \Delta^{{\rm nloc}}_{13}$ are
addressed.  It has been demonstrated in \cite{Karlin98} that the
first local correction to the $13$-moment approximation results in
Grad's equations corrected in the sense that the transport
coefficients become the \textit{exact} Chapman-Enskog coefficients
(not the first Sonine polynomial approximation, as in the original
Grad equations).  Whether or not the local correction is
spectacular in the context of the single-component gas with
traditional collision models like hard spheres (the first Sonine
polynomial approximation is "good" already...), the clear
distinction between local and nonlocal corrections is crucial. For
example, with this distinction it is possible to extend the method
of invariant manifolds to driven systems (see the derivation of
the Oldroyd constitutive equations from polymer kinetic theory
\cite{ZKD00}).

Now, let us turn our attention to the nonlocal piece of the defect of
invariance.  It can be demonstrated that
$\Delta^{{\rm nloc}}_{13}$ \textit{contains no terms with gradients of
  neither of the five hydrodynamic fields}, the only gradient
contributing to $\Delta^{{\rm nloc}}_{13}$ are of the stress tensor
and of the heat flux.  In the linear approximation near the global
equilibrium $F_0$ \cite{Karlin98},

\begin{equation}
\label{nonloc}
\Delta_{13}^{{\rm
nloc}}=-v_T^0F_{0}(\Pi_{1|krs}\partial_k \sigma_{rs}
+\Pi_{2|ik}\overline{\partial_k q_i} +\Pi_{3}\partial_k q_k)
\end{equation}

where $\partial_i =\partial/\partial x_i$, and $\Pi$ are velocity
polynomials:
\begin{eqnarray*}
\Pi_{1|krs}&=&c_k\left[c_rc_s-(1/3)\delta_{rs}c^2\right]-(2/5)
\delta_{ks}c_rc^2,
\\\nonumber
\Pi_{2|ik}&=&(4/5)\left[c^2-(7/2)\right]\left[c_ic_k-(1/3)
\delta_{ik}c^2\right],
\\\nonumber
\Pi_3&=&(4/5)\left[c^2-(5/2)\right]\left[c^2-(3/2)\right]-c^2.
\end{eqnarray*}

The absence of the gradients of the hydrodynamic fields in the
nonlocal defect reveals some important information: The invariance
correction to the $13$-moment approximation differs from the
higher-order corrections to the hydrodynamic equations.  For
example, since the linearized hydrodynamic equations following
from the $13$-moment Grad (uncorrected) equations to second order
in Knudsen number are the Burnett equations for Maxwell molecules
(see, e.\ g.,\ comparison of corresponding dispersion relations in
\cite{GK91a}), the same is true also for the corrected Grad
equations, as explicitly verified by Struchtrup and Torillhon
\cite{Struchtrup2002a}.

After this qualitative analysis of the defect of the invariance of
Grad's approximation, let us finish setting up the invariance equation
of the first iteration formally.  With the replacement (\ref{replace})
in the equation (\ref{itqe}), and using $P_0L_{f_0}=0$, we have,

\begin{equation}
\label{it13} (1-\Pi_{13})L_{f_0}\delta
f+(P_{13}-1)(\vv-\uu)\cdot\grad\delta
  f+\Delta_{13}^{{\rm loc}}+ \Delta^{{\rm nloc}}_{13}=0.
\end{equation}

In principle, this equation can be studied in the same spirit as the
equation of the first iteration to local Maxwellians, that is, without
introducing small parameters.  However, it is much more instructive to
consider the collision-dominant case, introducing the scaling
(\ref{kn}),

\begin{equation}
\label{it13kn} (1-\Pi_{13})\frac{1}{\epsilon}L_{f_0}\delta
f+(P_{13}-1)(\vv-\uu)\cdot\grad\delta
  f+\frac{1}{\epsilon}\Delta_{13}^{{\rm loc}}+ \Delta^{{\rm nloc}}_{13}=0.
\end{equation}

The correction $\delta f$ to first \textit{two orders}, $\delta
f\simeq \delta f^{(0)}+\epsilon \delta f^{(1)}$ is found from
equations:

\begin{eqnarray}
\label{loc} (1-\Pi_{13})L_{f_0}\delta
f^{(0)}=-\Delta_{13}^{{\rm loc}},\\
(1-\Pi_{13})L_{f_0}\delta f^{(1)}=-\Delta^{{\rm nloc}}_{13}.
\end{eqnarray}
These equations have to be solved subject to the additional
conditions, $P_{13}\delta f^{(0)}=0$, and $P_{13}\delta f^{(1)}=0$,
respectively.

The first equation (\ref{loc}) is responsible for the local
correction, as expected.  It's significance was discussed above.  For
the following, we assume no local correction is needed, that is,
either we assume BGK or Maxwell molecules (then $\delta f^{(0)}=0$
rigorously), or that the Grad's approximation is a reasonably good
approximation for the eigenvectors of $L_{f_0}$.

\begin{equation}\label{pheno}
 \delta f^{(1)}=\frac{1}{\tau}\Delta^{{\rm nloc}}_{13}.
\end{equation}

As it was emphasized in Ref.\ \cite{Karlin98}, {\it the invariance
correction to the $13$-moment Grad's approximation is related to
the $13$-moment Grad equations entirely in the same way as the
Navier-Stokes equations are related to the Euler equations}.
Roughly speaking, it uses the same amount of the Boltzmann
collision integral as the classical first-order equation of the
Chapman-Enskog method.  For that reason, this is the distinguished
case among other possible applications of MIM to improve on moment
approximations.

\subsection*{Strongly nonlinear invariance corrections}

As we have seen it in the previous section, the invariance
correction to the quasi-linear quasi-equilibria (Grad's moment
approximations) is distinguished by the fact that we can compute
it with the usual linearized collision integral in the local
equilibrium. Then the resulting linear integral equations have the
same structure as in the classical Chapman-Enskog method,
$L_{f_0}\varphi=\Delta$ (albeit with a different right hand side
$\Delta$).  The operator $L_{f_0}$ is self-adjoint in the scalar
product generated by the second differential of the entropy in the
local equilibrium, and thus it is simple to solve.  This is not
the case when the manifolds we want to correct contain pieces well
beyond the vicinity of the local equilibrium, for example, general
quasi-equilibria.  In these cases,
 the linearized collision operator $L_{f(M)}$ is not self-adjoint anymore.
In such cases, it was suggested to use the {\it symmetric linearization}
in order to establish dynamic corrections in highly nonequilibrium
situations. The symmetric
linearization of the Boltzmann collision integral in the state $f$ has the form,

\begin{equation}
L_{f}^{\textrm{sym}}\delta f= \int w \frac
{f^{\prime}f_1^{\prime}+ff_1}{2} \left[\frac{\delta
f^\prime}{f^{\prime}}+ \frac{\delta f_1^{\prime}}{f_1^{\prime}}
-\frac{\delta f_1}{f_1}-\frac{\delta f}{f}\right]
d{\vv}_1^{\prime}d{\vv}^{\prime}d{\vv}_1.\label{SYM}
\end{equation}

Note that $L_{f}^{\rm sym}\to L_{f_0}$ if the state $f$ tends to the local
Maxwellian $f_0$ (the consequence of the detail balance,
$f'f_1'=ff_1$ in the local equilibrium). Operator $L_{f}^{\textrm{sym}}$ enjoys
the familiar properties of the usual linearized collision integral.
Let us introduce notation for the {\it entropic scalar product} in the state $f$:
For two distribution functions $g_1$ and $g_2$,
\begin{equation}
\label{entprod}
\langle g_1|g_2\rangle_f=\int \frac{g_1(\vv)g_2(\vv)}{f(\vv)}d\vv.
\end{equation}
The following three properties of the operator $L_f^{\rm sym}$ are
immediate consequence of the definitions (\ref{SYM}) and (\ref{entprod}):

\begin{enumerate}
\item[(i) ]
$\langle g_1|L_f^{\rm sym}|g_2\rangle_f=
\langle g_2|L_f^{\rm sym}|g_1\rangle_f$ (symmetry);
\item[(ii) ]
$\langle g|L_f^{\rm sym}|g\rangle_f\leq 0 $ (local entropy production inequality);
\item[(iii)]
$L_f^{\rm sym}g=0$ if $g/f\in \mbox{Lin}\{1,{\bf v},v^2\}$ (conservation laws).
\end{enumerate}

Using the symmetric linearization, we see the equation for the
invariance correction for a general quasi-equilibrium $f(M)$
becomes
\begin{equation}
 (1-P)L^{\rm sym}_{f(M)}\delta f+(P-1)(\vv-\uu)\cdot\grad\delta
  f+\Delta(M)=0.
\end{equation}
The only difference with the equation (\ref{itqe}) is in the
replacement of the linearized collision integral $L_{f(M)}$ with
the symmetric linearized collision integral $L_{f(M)}^{\rm sym}$.
This difference is crucial though: When using symmetric operator
(\ref{SYM}), we get back all the familiar tools for solving
integral equations (Fredholm alternative in the
collision-dominated case \cite{Chapman}, the parametrix expansion
without such domination \cite{GK94inv},  and like).  All this is
impossible with the plain linearized operator $L_{f(M)}$, for
example, even the null-space of $L_{f(M)}$ is not known in
general.

The symmetric iteration was tested in the case of
finite-dimensional kinetic systems of chemical kinetics
\cite{GKZD2000,GK2002} (see, in particular, its recent application
to construction of grid representations of invariant manifolds
\cite{GKZ2003}).  Results of convergence of symmetric iterations
to slow nonlinear manifolds are quite encouraging. At the time of
this writing, symmetric iteration remains almost entirely
unexplored for the Boltzmann equation.

\subsection*{Invariance principle in the moment representation}

The aforementioned computations of various quasi-equilibria,
moment approximations and their invariance corrections were all
done in the setting of the kinetic Boltzmann equation and
distribution functions. Invariance corrections can also be studied
in a simplified setting: Consider a closed system for $n=k+m$
moments and reduce it to a closed system for $k$ of them.  Such a
simplification (with respect to the full kinetic theory) makes
sense especially if one wants to get a basic qualitative
understanding about the form of reduced description in terms of
$k$ moments.

This problem was studied to some very detailed extend, and well
beyond the usual first-order Knudsen number corrections, for the
case of hydrodynamics from $10$- and $13$-moment Grad equations
beginning with the paper \cite{GK91} on a \textit{partial
summation of the
  Chapman-Enskog expansion to all the orders in Knudsen number}, and
on the \textit{exact summation} of the expansion \cite{GK96a}.
Some of these studies were recently summarized in
\cite{Karlin2002a} with the emphasis of the iteration method for
solving the invariance equation, and the interested reader is
directed to that paper. In spite of a seemingly drastic
simplification with respect to the "true" kinetic theory, results
are sometimes surprisingly robust.  For example, the leading
invariance correction to the nonlinear longitudinal viscosity in
so-called homoenergetic flow found from the $10$-moment equations
\cite{KDN97} is \textit{exactly} the one found independently from
the exact solution to the model Boltzmann equation
\cite{Santos2000}. New interesting results on summation of the
Chapman-Enskog expansion for the Boltzmann equation were obtained
recently by Slemrod \cite{Slemrod98}.

The framework of a larger moment system used to obtain the
invariance correction to a smaller moment system appears in the
recent work of Struchtrup and Torrilhon \cite{Struchtrup2002a}.
The difference with the derivation from the Boltzmann equation is
basically the absence of the local correction. The study
\cite{Struchtrup2002a} demonstrated a set of  advantages of these
equations above the Grad's system, most importantly, the improved
shock wave structure.

\section{Quasi-equilibrium kinetic models}

The invariance corrections explore more of the phase space than
initially assumed by making a Grad approximation with a given number
of variables.  By measuring the defect of invariance, we realize in
which direction the quasi-equilibrium manifold should be improved in
order to take into account fast motion towards it.  There is another
useful way to explore fast motions: To lift the dynamics from the
manifold to a dynamics in the full space by means of a kinetic model.

We recall that lifting the Euler dynamics which takes place on the local Maxwell
manifold to a kinetics in the whole phase space  is done
by the very useful Bhatnagar-Gross-Krook model (BGK),
\begin{equation}
\label{BGK}
\partial_t f+\vv\cdot\nabla f=-\frac{1}{\tau}(f-f_0(f)),
\end{equation}
where $\tau>0$ is the relaxation time, and
$f_0(f)$ is a map $f\to f_0$ established by local conservation laws:
\begin{equation}
\label{consistency}
\int\{1,\vv,v^2\}(f-f_0(f))d\vv=0.
\end{equation}
The right hand side of Eq.\ (\ref{BGK}),
\begin{equation}
\label{BGKcollint} Q_{\rm BGK}=-\frac{1}{\tau}(f-f_0(f)),
\end{equation}
is called the BGK collision integral. Proof of the $H$-theorem for the
BGK kinetic equation does not rely anymore on the
microscopic reversibility (as in the Boltzmann case), instead, it is
a simple consequence of convexity of the $H$-function, and
of the property of the map (\ref{consistency}):
\begin{eqnarray}
\label{sig}
\sigma&=&-\frac{1}{\tau}
\int \ln f(f-f_0(f))d\vv\nonumber\\&=&
-\frac{1}{\tau}\int\ln\left(\frac{f}{f_0(f)}\right)(f-f_0(f))d\vv\le 0.
\end{eqnarray}

Now, how to lift general quasi-equilibria (and, consequently, also the Grad approximations)
to a kinetic model? The answer to this question was given in the Ref.\
\cite{GK94bgk}. The kinetic model for a quasi-equilibrium approximation
$f(M)$ has the form:
\begin{equation}
\partial_t f+\vv\cdot\nabla f=-\frac{1}{\tau}(f-f(M(f)))
+Q(f(M(f)),f(M(f))). \label{gBGK}
\end{equation}
Here $f(M(f))$ is the natural map $f\to f(M)$,
\begin{equation}
\label{gmap}
\int m_k(f-f(M(f)))d\vv=0,\ k=1,\dots,n,
\end{equation}
and thus the first term in the right hand side of equation (\ref{gBGK}) is just
BGK-like, whereas the second term, function $Q(f(M(f)),f(M(f)))$ is
the true (Boltzmann) collision integral, {\it evaluated on the quasi-equilibrium
manifold}. The latter is crucial: Unlike the true Boltzmann collision integral $Q(f,f)$ which
can take values in the entire phase space of distribution function,
 $Q(f(M(f)),f(M(f)))$ is allowed to take values only on a relatively thin
subset known a priori, and can be thus {\it pre-computed} to the explicit function
of $M$ and $\vv$ (see Ref.\ \cite{GK94bgk} for examples).
If the quasi-equilibrium $f(M)$ consists only of the local Maxwellians,
then  $Q(f(M(f)),f(M(f)))$ equals to zero, and we get back the BGK-model.
In all other cases, the second term in the kinetic model (\ref{gBGK}) is
essential: If it is omitted in equation (\ref{gBGK}) then the
zero of the resulting collision integral is the whole quasi-equilibrium
manifold $f(M)$, and not its local Maxwellian submanifold, unlike the case
of the Boltzmann collision integral.

The $H$-theorem for kinetic models (\ref{gBGK}) has the following structure
\cite{GK94bgk}: Let us compute $\sigma$ (\ref{sig}):
\begin{eqnarray}
\sigma&=&\sigma_{\rm BGK}+\sigma_Q,\nonumber\\
\sigma_{\rm BGK}&=&-\frac{1}{\tau}\int \ln(f) (f-f(M(f))d\vv,\nonumber\\
\sigma_Q&=&\int \ln(f)Q(f(M(f)),f(M(f)))d\vv
\end{eqnarray}
Function  $\sigma_{\rm BGK}$ is the contribution from the BGK-like
term in equation (\ref{gBGK}), and it is  always  non-positive,
again due to the property of the map $f\to f(M)$ (\ref{gmap}). The
second contribution, $\sigma_Q$ may be not sign-definite if $f$ is
taken far away from the quasi-equilibrium. However, one proves
\cite{GK94bgk} that there always exists a non-empty neighborhood
of the quasi-equilibrium manifold, where $\sigma_Q\le 0$ (this is
almost obvious: {\it On} the quasi-equilibrium manifold
$\sigma_Q(f(M))$ is the  entropy production due to the true
Boltzmann collision integral). Thus, if the relaxation towards
quasi-equilibrium states is fast enough ($\tau$ is sufficiently
close to zero), the net entropy production inequality holds,
$\sigma=\sigma_{\rm BGK}+\sigma_Q\le 0$.

Further simplification of the models (\ref{gBGK}) are possible.
Let us mention here
two of them: First, instead of function $Q(f(M(f)),f(M(f)))$, we can use
\begin{eqnarray}
PQ(f(M(f)),f(M(f)))=\sum_k\frac{\partial f(M)}{\partial M_k}
\bigg|_{M=M(f)}R_k(M(f)),
\nonumber\\
R_k(M(f))=\int m_k Q(f(M(f)),f(M(f))) d\vv.
\end{eqnarray}
That is, instead of the true collision integral $Q$, we take only
its quasi-equilibrium
projection, $PQ$. The simplification here is that the velocity dependence
is now accumulated only in the quasi-equilibrium distribution, and not
in the function $Q(f(M(f)),f(M(f)))$:
\begin{equation}
\partial_t f+\vv\cdot\nabla f=-\frac{1}{\tau}(f-f(M(f)))
+ \sum_k\frac{\partial f(M)}{\partial
M_k}\bigg|_{M=M(f)}R_k(M(f)).
\end{equation}
Second, and probably the last simplification occurs if one uses
the BGK collision integral $Q_{\rm BGK}$ (\ref{BGKcollint}), with
a different relaxation time, say $\theta$, instead of the
Boltzmann collision integral:
\begin{equation}
\partial_t f+\vv\cdot\nabla f=-\frac{1}{\tau}(f-f(M(f)))
-\frac{1}{\theta}\sum_k\frac{\partial f(M)}{\partial
M_k}\bigg|_{M=M(f)} (M_k(f)-M_k^{(0)}(f(M))).
\end{equation}
Here $M_k^{(0)}$ denotes the $k$-order moment of the local Maxwellian.

As a final comment here, the family of the kinetic models reviewed in this
section use the {\it natural map} $f\to f(M)$  (\ref{gmap}) of the quasi-equilibrium
approximations. Different  maps $f \to f(M)$ which do not obey (\ref{gmap}) were used recently
to establish BGK-type models for various quasi-equilibrium approximations
\cite{Perthame2002}.

\section{Lattice Boltzmann and other minimal kinetic models}
\label{LBsection}

The past decade has witnessed a rapid development of {\it minimal
kinetic models} for numerical simulation of complex macroscopic
systems. The lattice Boltzmann method is particularly valuable
minimal extension of the Navier-Stokes equation finding
increasingly more applications in computational fluid dynamics.
Some relation of the lattice Boltzmann method to Grad's method was
indicated in \cite{He98}: Once the Grad method is supplemented by
the Gauss-Hermite quadrature in the velocity space, the moment
system can be rewritten in the form of a discrete-velocity model,
that is, it becomes amenable to effective numerical
implementation. Recently, a quasi-equilibrium version of this
construction was established, in which the quadrature is done not
on the distribution function but on the entropy functional
\cite{AK2002,AK2002a}. Quite remarkably, the quasi-equilibrium
perspective on the lattice Boltzmann method results in its
refinement known as the {\it entropic lattice Boltzmann method}
\cite{KGSB98, KFOe99,Boghosian01}. Here we review the entropic
lattice Boltzmann method (ELBM) for hydrodynamics.

We start with a generic discrete velocity kinetic model. Let
$f_i(\xx,t)$ be populations of the $D$-dimensional discrete
velocities $\vel_i$, $i=1,\dots,n_{\rm d}$, at position $\xx$ and
time $t$. The hydrodynamic fields are the linear functions of the
populations, namely
\begin{equation}\label{fielshyd}
\sum_{i=1}^{n_{\rm d}} \{ 1,\, \vel_{i},\, v_i^2 \} f_i =\{\rho,\,
\rho \uu, \, \rho DT+ \rho u^2 \},
\end{equation}
where $\rho$ is the mass density of the fluid, $\rho \uu$ is the
$D$-dimensional momentum density vector, and $e=\rho D T+ \rho
u^2$ is the energy density. In the case of isothermal simulations,
the set of independent hydrodynamic fields contains only the mass
and momentum densities. It is convenient to introduce $n_{\rm
d}$-dimensional population vectors $\ff$, and the standard scalar
product, $(\ff|\gggg)=\sum_{i=1}^{{n_{\rm d}}}x_iy_i$. For
example, for almost-incompressible hydrodynamics (leaving out the
energy conservation), the locally conserved density and momentum
density fields are written as
\begin{equation}\label{Hyd}
( \bONE|\ff)=\rho,\ ( \vel_{\alpha}|\ff)=\rho u_{\alpha}.
\end{equation}
Here $\bONE=\{ 1\}_{i=1}^{n_{\rm d}}$,
$\vel_{\alpha}=\{v_{i\alpha}\}_{i=1}^{n_{\rm d}}$, and
$\alpha=1,\dots,D$, where $D$ is the spatial dimension.

The construction of the kinetic simulation scheme begins with
finding a convex function of populations $H$ (entropy function),
which satisfies the following condition: If $\ff^{\rm
eq}(\rho,\uu)$ (local equilibrium) minimizes $H$ subject to the
hydrodynamic constraints (equations (\ref{fielshyd}) or
(\ref{Hyd})), then $\ff^{\rm eq}$ also satisfies certain
restrictions on the higher-order moments. For example, the
equilibrium stress tensor must respect the Galilean invariance,
\begin{equation}
\label{GI}
  \sum_{i = 1}^{{n_{\rm d}}} v_{i\alpha}v_{i\beta}f^{\rm eq}_i(\rho,\uu)=
\rho c_{\rm s}^2\delta_{\alpha\beta}+\rho u_{\alpha}u_{\beta}.
\end{equation}
The corresponding entropy functions for the isothermal and the
thermal models were found in
\cite{KFOe99,AK2002a,AK2002b,AKOe2003}, and are given below (see
section \ref{Hfunc} and Table \ref{Tab: DiscV}). For the time
being, assume that the convex function $H$ is given.

The next step is to obtain the set of kinetic equations,
\begin{equation}
\label{LBMcont} \partial_t f_i+v_{i\alpha}\partial_{\alpha}f_i=
\Delta_i.
\end{equation}
Let $\mm_1,\dots,\mm_{n_{\rm c}}$ be the ${n_{\rm d}}$-dimensional
vectors of locally conserved fields, $M_i=( \mm_i|\ff)$,
$i=1,\dots,{n_{\rm c}}$,
 $n_{\rm c}<{n_{\rm d}}$. The ${n_{\rm d}}$-dimensional vector function $\bDelta$
(collision integral), must satisfy the conditions:
\[ ( \mm_i|\bDelta)=0,\ i=1,\dots,n_{\rm c}\
(\rm{local\ conservation\ laws}),\]
\[ (\bnabla H|\bDelta)\le 0\
(\rm{entropy\ production\ inequality}),\] where $\bnabla H$ is the
row-vector of partial derivatives $\partial H/\partial f_i$.
Moreover, the local equilibrium vector $\ff^{\rm eq}$ must be the
only zero point of $\bDelta$, that is, $\bDelta(\ff^{\rm
eq})=\mathbf{0}$, and, finally, $\ff^{\rm eq}$ must be the only
zero point of the local entropy production, $\sigma(\ff^{\rm
eq})=0$. Collision integrals which satisfies all these
requirements are called admissible. Let us discuss several
possibilities of constructing admissible collision integrals.

\subsection*{BGK model}

Suppose that the entropy function $H$ is known. If, in addition,
the local equilibrium is also known as an explicit function of the
locally conserved variables (or some reliable approximation of
this function is known), the simplest option is to use the
Bhatnagar-Gross-Krook (BGK) model. In the case of isothermal
hydrodynamics, for example, we write
\begin{equation}\label{BGKcoll}
\bDelta=-\frac{1}{\tau}(\ff-\ff^{\rm eq}(\rho(\ff), \uu(\ff))).
\end{equation}
The BGK collision operator is sufficient for many applications.
However, it becomes advantageous only if the local equilibrium is
known in a closed form. Unfortunately, often only the entropy
function is known but not its minimizer. For these cases one
should construct collision integrals  based solely on the
knowledge of the entropy function. We present here two particular
realizations of the collision integral based on the knowledge of
the entropy function only.

\subsection*{Quasi-chemical model}

For a generic case of $n_{\rm c}$ locally conserved fields, let
$\gggg_s$, $s=1,\dots, {n_{\rm d}}-{n_{\rm c}}$, be a basis of the
subspace orthogonal (in the standard scalar product) to the
vectors of the conservation laws. For each vector $\gggg_{s}$, we
define a decomposition $\gggg_{s}=\gggg_{s}^+-\gggg_{s}^-$, where
all components of vectors $\gggg_{s}^{\pm}$ are nonnegative, and
if $g_{si}^{\pm}\ne0$, then $g_{si}^{\mp}=0$. Let us consider the
collision integral of the form:
\begin{equation}
\label{lbMDD} \bDelta=\sum_{s=1}^{{n_{\rm d}}-{n_{\rm
c}}}w_{s}\gggg_{s} \left\{ \exp\left((\bnabla
H|\gggg_{s}^-)\right)- \exp\left((\bnabla H|\gggg_{s}^+)\right)
\right\}.
\end{equation}
Here $w_{s}>0$. By construction, the collision integral
(\ref{lbMDD}) is admissible. If the entropy function is
Boltzmann--like, and the components of the vectors $\gggg_{s}$ are
integers, the collision integral assumes the familiar
Boltzmann--like form.

\subsection*{Single relaxation time gradient model}

The BGK collision integral (\ref{BGKcoll}) has the following
important property: the linearization of the operator
(\ref{BGKcoll}) around the local equilibrium point has a very
simple spectrum $\{0,-1/\tau\}$, where $0$ is the ${n_{\rm
c}}$-times degenerate eigenvalue corresponding to the conservation
laws, while the non-zero eigenvalue corresponds to the rest of the
(kinetic) eigenvectors. Nonlinear collision operators which have
this property of their linearizations at equilibrium are called
single relaxation time models (SRTM). They play an important role
in modelling because they allow for the simplest identification of
transport coefficients.

The SRTM, based on the given entropy function $H$, is constructed
as follows (single relaxation time gradient model, SRTGM). For the
system with ${n_{\rm c}}$ local conservation laws, let $\ee_s$,
$s=1,\dots,{n_{\rm d}}-{n_{\rm c}}$, be an orthonormal basis in
the kinetic subspace, $(\mm_i|\ee_s)=0$, and
$(\ee_s|\ee_p)=\delta_{sp}$. Then the single relaxation time
gradient model is
\begin{equation}
\label{family} \bDelta=-\frac{1}{\tau}\sum_{s,p=1}^{{n_{\rm
d}}-{n_{\rm c}}}\ee_{s} K_{sp}(\ff) (\ee_{p}|\bnabla H),
\end{equation}
where $K_{sp}$ are elements of a positive definite $({n_{\rm
d}}-{n_{\rm c}})\times({n_{\rm d}}-{n_{\rm c}})$ matrix $\KK$,
\begin{eqnarray}
\label{SRTM} \KK(\ff)&=&\CC^{-1}(\ff),\\\nonumber
C_{sp}(\ff)&=&(\ee_s|\bnabla\bnabla H(\ff) |\ee_{p}).
\end{eqnarray}
Here, $\bnabla\bnabla H(\ff)$ is the ${n_{\rm d}}\times {n_{\rm
d}}$ matrix of second derivatives, $\partial^{2}H/\partial
f_i\partial f_j$. Linearization of the collision integral at
equilibrium has the form,
\begin{equation}
\label{P} \LL=-\frac{1}{\tau}\sum_{s=1}^{{n_{\rm d}}-{n_{\rm
c}}}\ee_s\ee_s,
\end{equation}
which is obviously single relaxation time. Use of the SRTGM
instead of the BGK model results in the same hydrodynamics even
when the local equilibrium is not known in a closed form. Further
details of this model and its numerical implementation can be
found in Ref. \cite{AK2002a}.


It is pertinent to our discussion to explain the term ``gradient"
appearing in the name SRTGM.
In Euclidean spaces with the given scalar product, we often
identify the differential of a function $f(x)$ with its gradient:
in the orthogonal coordinate system   $({\rm grad} f(x))_i =
\partial f(x) / \partial x_i $. However, when dealing with a more general setting, one can run into problems while making
sense out of such a definition. What to do, if there is no
distinguished scalar product, no preselected orthogonality?

For a given scalar product $\langle \: | \: \rangle$ the gradient
${\rm grad}_x f(x)$ of a function $f(x)$ at a point $x$ is such a
vector $g$ that $\langle g | y\rangle = D_xf (y)$ for any vector
$y$, where $D_xf$ is the differential of function $f$ at a point
$x$. The differential of function $f$ is the linear functional
that provides the best linear approximation near the given point.

In order to transform a vector into a linear functional one needs
a \index{Pairing}{\it pairing}, that means a bilinear form
$\langle \: | \: \rangle$. This pairing transforms vector $g$ into
linear functional $\langle g|$:
 $\langle g|(x)=\langle g|x\rangle$. Any twice differentiable function $f(x)$ generates a field of pairings:
at any point $x$ there exists a second differential of $f$, a
quadratic form $(D^2_x f)(\Delta x,\Delta x)$. For a convex
function these forms are positively definite, and we return to the
concept of scalar product. Let us calculate a gradient of $f$
using this scalar product. In coordinate representation the
identity $\langle{\rm grad} f(x) \mid y \rangle_x = (D_x f) (y)$
(for any vector $y$) has a form
\begin{equation}\label{New0}
\sum_{i,j}({\rm grad} f(x))_i \frac {\partial^2 f}{\partial x_i
\partial x_j}y_j= \sum_{i} \frac {\partial f}{\partial x_j}y_j,
\end{equation}
hence,
\begin{equation}\label{New}
({\rm grad} f(x))_i = \sum_{j} (D^2_x f)^{-1}_{ij}\frac {\partial
f}{\partial x_j}.
\end{equation}
As we can see, this ${\rm grad} f(x)$ is the \index{Newtonian
direction} {\it Newtonian direction}, and with this gradient the
method of steepest descent transforms into the Newton method of
optimization.

Entropy is the concave function, and we define the entropic scalar
product through negative second differential of entropy. Let us
define the gradient of entropy by means of this scalar product:
$\langle{\rm grad}_{x} S|z\rangle_{x} = (D_{x}S)(z)$. The
\index{Entropic!gradient system}{\it entropic gradient system} is
\begin{equation}\label{entgrad}
\frac{ \,\D x}{ \,\D t}=\varphi({x}){\rm grad}_{x} S,
\end{equation}
where $\varphi({x})>0$ is a positive kinetic multiplier.
The entropic gradient models (\ref{entgrad}) possesses all the
required properties (if the entropy Hessian is sufficiently
simple). In many cases it is simpler than the BGK model, because
the gradient model is {\it local} in the sense that it uses only
the entropy function and its derivatives at a current state, and
it is not necessary to compute the equilibrium (or
quasi-equilibrium for quasi-equilibrium models).
The entropic gradient model has a one-point relaxation spectrum,
because near the equilibrium $x^{\rm eq}$ the gradient vector
field (\ref{entgrad}) has  an extremely simple linear
approximation: $ \,\D (\Delta x)/ \,\D t=-\varphi({x^{\rm
eq}})\Delta x$. It corresponds to a well-known fact that the
Newton method minimizes a positively defined quadratic form in one
step. The SRTGM discrete velocity model (\ref{family}),
(\ref{SRTM}) is a particular realization of this construction when
the local conservation laws are projected out.

\subsubsection{$H$-functions of minimal kinetic models} \label{Hfunc}

The Boltzmann entropy function written in terms of the
one-particle distribution function $f({\xx}, \vel)$ is $H=\int
f\ln f \,d \vel$, where $\vel$ is the continuous velocity. Close
to the global (reference) equilibrium, this integral can be
approximated by using the Gauss--Hermite quadrature with the
weight
\[ W= (2\, \pi \, T_0)^{(D/2)}
\exp(-v^2/(2\,T_0)).\] Here $D$ is the spatial dimension, $T_0$ is
the reference temperature, while the particles mass and
Boltzmann's constant $k_{\rm B}$ are set equal to one. This gives
the entropy functions of the discrete-velocity models
\cite{KFOe99,AK2002b,AKOe2003},
\begin{equation}
\label{app:H} H=\sum_{i=1}^{{n_{\rm d}}}
f_{i}\ln\left(\frac{f_{i}}{w_i} \right).
\end{equation}
Here, $w_i$ is the weight associated with the $i$-th discrete
velocity $\vel_i$ (zeroes of the Hermite polynomials).  The
discrete-velocity distribution functions (populations)
$f_i({\xx})$ are related to the values of the continuous
distribution function at the nodes of the quadrature by the
formula,
\[f_i({\xx})=w_i(2\, \pi \, T_0)^{(D/2)}
\exp(v^2_i/(2\,T_0))f({\xx}, {\vel}_i).\]  The  entropy functions
(\ref{app:H}) for various $\{w_i,\vel_i\}$ are the only input
needed for the construction of minimal kinetic models.

With the increase of the order of the Hermite polynomials used in
evaluation of the quadrature (\ref{app:H}), a better approximation
to the hydrodynamics is obtained. The first few models of this
sequence are represented in Table \ref{Tab: DiscV}.

\begin{table}[t]
  \caption{\label{Tab: DiscV} Minimal kinetic models \cite{AKOe2003}.
Column 1: Order of Hermite velocity polynomial used to evaluate
  the Gauss-Hermite quadrature;
Column 2: Locally conserved (hydrodynamic) fields; Column 3:
Discrete velocities for $D=1$ (zeroes of the corresponding Hermite
polynomials). For $D>1$, discrete velocities are all possible
tensor products of the one-dimensional velocities in each
component direction; Column 4: Weights in the entropy formula
(\ref{app:H}), corresponding to the discrete velocities of the
Column 3. For $D>1$, the weights of the discrete velocities are
products of corresponding one-dimensional weights; Column 5:
Macroscopic equations for the fields of Column 2 recovered in the
hydrodynamic limit of the model.  }
   \bigskip
\begin{tabular}{|l|l|l|l|l|}
  \hline
  1. Order & 2. Fields    & 3. Velocities & 4. Weights & 5. Hydrodynamic limit \\
  \hline
  $2$ & $\rho$ & $ \sqrt{T_0}$ & $1/2$ & Diffusion   \\
      &        & $-\sqrt{T_0}$ & $1/2$ &             \\
  \hline
  $3$ & $\rho$, $\rho\uu$ & $0$                   & $2/3$ & Isothermal Navier--Stokes \\
      &                   & $\sqrt{3}\sqrt{T_0}$  & $1/6$ &
\\
      &                   & $-\sqrt{3}\sqrt{T_0}$ & $1/6$ &
  \\
  \hline
  $4$ &  $\rho$, $\rho\uu$, $e$ & $ \sqrt{3-\sqrt{6}}\sqrt{T_0}$ & $1/[4(3-\sqrt{6})]$ & Thermal Navier-Stokes \\
      &                         & $-\sqrt{3-\sqrt{6}}\sqrt{T_0}$ & $1/[4(3-\sqrt{6})]$ &                       \\
      &                         & $ \sqrt{3+\sqrt{6}}\sqrt{T_0}$ & $1/[4(3+\sqrt{6})]$ &                       \\
      &                         & $-\sqrt{3+\sqrt{6}}\sqrt{T_0}$ & $1/[4(3+\sqrt{6})]$ &                       \\
      \hline
\end{tabular}
\end{table}

\subsection*{Entropic lattice Boltzmann method}

\begin{figure}[t]
\centering{
\includegraphics[height=.30\textheight]{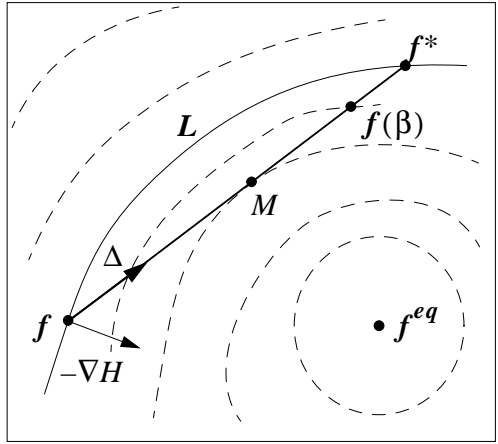}
\caption{\label{LBFig1} Entropic stabilization of the lattice
Boltzmann scheme with over-relaxation. Curves represent entropy
levels, surrounding the local equilibrium $\ff^{\rm eq}$. The
solid curve $L$ is the entropy level with the value
$H(\ff)=H(\ff^*)$, where $\ff$ is the initial, and $\ff^*$ is the
maximally over-relaxed population $\ff+\alpha\bDelta$ defined by
equation (\ref{HTdiscrete}). The vector $\bDelta$ represents the
collision integral, the sharp angle between $\bDelta$ and the
vector $-\bnabla H$ reflects the entropy production inequality.
The point $\MM$ is the state of minimum of the entropy function
$H$ on the line segment between $\ff$ and $\ff^*$. The result of
the collision update is represented by the point $\ff(\beta)$. The
choice of $\beta$ shown corresponds to the over-relaxation:
$H(\ff(\beta))>H(\MM)$ but $H(\ff(\beta))<H(\ff)$. The particular
case of the BGK collision (not shown) would be represented by a
vector $\bDelta_{\rm BGK}$, pointing from $\ff$ towards $\ff^{\rm
eq}$, in which case $\MM=\ff^{\rm eq}$. Figure from Ref.
\cite{KFOe99}.}}
\end{figure}

If the set of discrete velocities forms the links of a Bravais
lattice (with possibly several sub-lattices), then the
discretization of the discrete velocity kinetic equations in time
and space is particularly simple, and leads to the entropic
lattice Boltzmann scheme. This happens in the important case of
the isothermal hydrodynamics. The equation of the entropic lattice
Boltzmann scheme reads
\begin{equation}
\label{LBMdiscrete} f_i(\xx+\cc_i\delta t,t+\delta t)-f_i(\xx,t)=
\beta\alpha(\ff(\xx,t))\Delta_i(\ff(\xx,t)),
\end{equation}
where $\delta t$ is the discretization time step, and
$\beta\in[0,1]$ is a fixed parameter which matches the viscosity
coefficient in the long-time large-scale dynamics of the kinetic
scheme (\ref{LBMdiscrete}). The function $\alpha$ of the
population vector defines the maximal over-relaxation of the
scheme, and is found from the entropy condition,
\begin{equation}
\label{HTdiscrete}
H(\ff(\xx,t)+\alpha\bDelta(\ff(\xx,t))=H(\ff(\xx,t)).
\end{equation}
The nontrivial root of this equation is found for populations at
each lattice site. Equation (\ref{HTdiscrete}) ensures the
discrete-time $H$-theorem, and is required in order to stabilize
the scheme if the relaxation parameter $\beta$ is close to one.
The geometrical sense of the discrete-time $H$-theorem is
explained in Fig.\ \ref{LBFig1}. We note in passing that the
latter limit is of particular importance in the applications of
the entropic lattice Boltzmann method to hydrodynamics because it
corresponds to vanishing viscosity, and hence to numerically
stable simulations of  very high Reynolds number flows.


\subsection*{Entropic lattice BGK method (ELBGK)}

An important simplification occurs in the case of the isothermal
simulations when the entropy function is constructed using
third-order Hermite polynomials (see Table \ref{Tab: DiscV}): the
local equilibrium population vector can be obtained in closed form
\cite{AKOe2003}. This enables the simplest entropic scheme -- the
entropic lattice BGK model -- for simulations of isothermal
hydrodynamics. We present this model in dimensionless lattice
units.

Let $D$ be the spatial dimension. For $D=1$, the three discrete
velocities are
\begin{equation}\label{1Dvel}
\vel=\{-1, 0, 1\}.
\end{equation}
For $D>1$, the discrete velocities are tensor products of the
discrete velocities of these one-dimensional velocities. Thus, we
have the $9$-velocity model for $D=2$ and the $27$-velocity model
for $D=3$. The $H$ function is Boltzmann-like:
\begin{equation}
\label{app:H27} H=\sum_{i=1}^{3^D} f_{i}\ln\left(\frac{f_{i}}{w_i}
\right).
\end{equation}
The weights $w_i$ are associated with the corresponding discrete
velocity $\vel_i$. For $D=1$, the three-dimensional vector of the
weights corresponding to the velocities (\ref{1Dvel}) is
\begin{equation}\label{weights1D} \ww = \left
\{\frac{1}{6},\frac{2}{3}, \frac{1}{6} \right \}.
\end{equation}
For $D>1$, the weights are constructed by multiplying the weights
associated with each component direction.

The local equilibrium minimizes the $H$-function (\ref{app:H})
subject to the fixed density and momentum,
\begin{equation}\label{Hyd27}
\sum_{i=1}^{3^D}f_i=\rho,\ \sum_{i=1}^{3^D}f_iv_{i\alpha}=\rho
u_{\alpha},\ \alpha=1,\dots,D.
\end{equation}
The explicit solution to this minimization problem reads,
\begin{equation}
\label{TED}
 f^{\rm eq}_i=\rho w_i\prod_{\alpha=1}^{D}
\left(2 -\sqrt{1+ 3 {u_{\alpha}^2}} \right)
\left(\frac{2\,u_{\alpha} + \sqrt{1+ 3\,u_{\alpha}^2}}{1-u_\alpha}
\right)^{c_{i\alpha}}.
\end{equation}
Note that the exponent, $v_{i\alpha}$, in (\ref{TED}) takes the
values $\pm 1, \, \mbox{and}\,\,  0$ only, and the speed of sound,
$c_{\rm s}$, in this model is equal to $1/\sqrt{3}$. The
factorization of the local equilibrium (\ref{TED}) over spatial
components is quite remarkable, and resembles the familiar
property of the local Maxwellians.

The entropic lattice BGK model for the local equilibrium
(\ref{TED}) reads,
\begin{equation}
\label{LBGKdiscrete} f_i(\xx+\vel_i\delta t,t+\delta
t)-f_i(\xx,t)= -\beta\alpha(f_i(\xx,t)-f_i^{\rm
eq}(\rho(\ff(\xx,t)),\uu(\ff(\xx,t))).
\end{equation}
The parameter $\beta$ is related to the  relaxation time $\tau$ of
the BGK model (\ref{BGKcoll}) by the formula,
\begin{equation}\label{relationtau}
\beta=\frac{\delta t}{2\tau+\delta t},
\end{equation}
and the value of the over-relaxation parameter $\alpha$ is
computed at each lattice site from the entropy estimate,
\begin{equation}
H(\ff-\alpha(\ff-\ff^{\rm eq}(\ff)))=H(\ff).
\end{equation}
In the hydrodynamic limit, the model (\ref{LBGKdiscrete})
reconstructs the Navier-Stokes equations with the viscosity
\begin{equation}
\label{viscELBGK} \mu=\rho c_{\rm s}^2\tau=\rho c_{\rm s}^2\delta
t\left(\frac{1}{2\beta}-\frac{1}{2}\right).
\end{equation}
The zero-viscosity limit corresponds to $\beta\to 1$.

\subsection*{Wall boundary conditions}

The boundary (a solid wall) $\partial R$ is specified at any point
$ {\xx} \in\partial R $ by the inward unit normal $\bm{e}$, the
wall temperature $T_{\rm wall }$, and the wall velocity
${\uu}_{\rm wall }$. The simplest boundary condition for the
minimal kinetic models presented above is obtained upon evaluation
of the diffusive wall boundary condition for the Boltzmann
equation with \cite{Cercignani75} the help of the Gauss-Hermite
quadrature. Details can be found in  \cite{AK2002b,kirp}. We here
write out the final expression for the diffusive wall boundary
condition:
\begin{equation}
\label{DBC}
  f_i
=  \frac{ \sum_{\mbox{ {\scriptsize $\bxi$}}_{i'} \cdot
\mbox{{\scriptsize $\nn$ }} < 0  } |(\bxi_{i'}\cdot{\nn})|
f_{i^{\prime}} }{ \sum_{\mbox{{\scriptsize $\bxi$}}_{i'} \cdot
\mbox{{\scriptsize $\nn$ }} < 0 } |(\bxi_{i'}\cdot{\nn})|
f_{i^{\prime}}^{\rm eq}(\rho_{\rm wall },\uu_{\rm wall })}f^{\rm
eq}_i(\rho_{\rm wall },\uu_{\rm wall }), \qquad ( \bxi_i\cdot
{\nn}
> 0).
\end{equation}
Here $\bxi_i$ is the discrete velocity  in the wall reference
frame, $ \bxi_i={\vel}_i - {\uu}_{\rm wall}$.

\subsection*{Numerical illustrations of the ELBGK}

\begin{figure}[t]
\centering{
\includegraphics[scale=0.5]{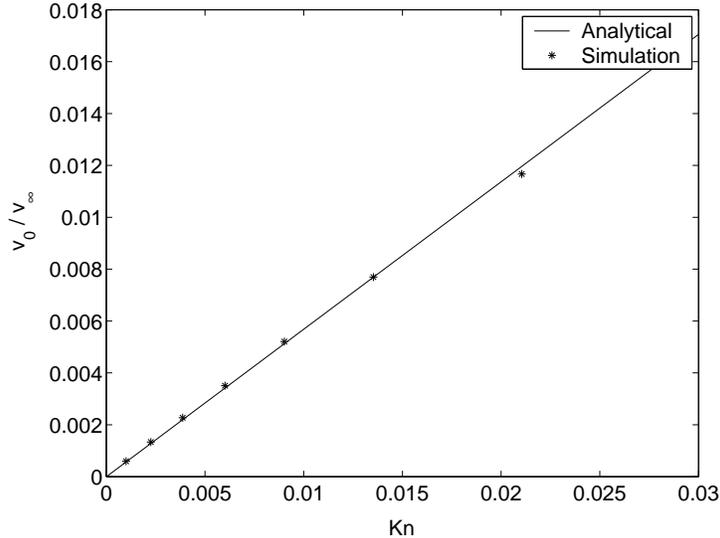}
\caption{ \label{LBFig2}  Relative slip at the wall in the
simulation of the Kramers problem for shear rate   $a = 0.001 $,
box length $L=32$, $v_\infty = a \times L = 0.032$. Figure from
Ref.\ \cite{AK2002b}, computed by S. Ansumali. }}
\end{figure}

The  Kramers  problem \cite{Cercignani75} is a limiting case of
the plane Couette flow, where one of the plates is moved to
infinity, while keeping a fixed shear rate. The analytical
solution for the \index{Slip-velocity}
slip-velocity at the wall
calculated for the linearized BGK collision model
\cite{Cercignani75}  with the simulation of the entropic lattice
BGK model are compared  in Fig. \ref{LBFig2}. This shows that one
important feature of original Boltzmann equation, the Knudsen
number dependent slip at the wall is retained in the present
model.

\begin{figure}[t]
\centering{
\includegraphics[scale=0.5]{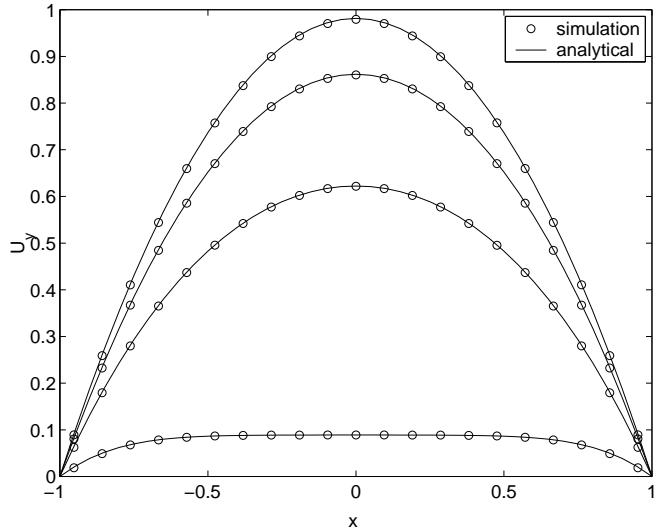}
\caption{\label{LBFig3} Development of the velocity profile in the
Poiseuille flow. Reduced velocity $U_y(x) = u_y/u_{y_{\rm max}}$
is shown versus the reduced coordinate across the channel $x$.
Solid line: Analytical solution. Different lines correspond to
different instants of the reduced time, increasing from bottom to
top. Symbol: simulation with the ELBGK algorithm. Parameters used
are: viscosity $\mu= 5.0015 \times 10^{-5}$  ($\beta = 0.9997$),
steady state maximal velocity
  $u_{y_{\rm max}} = 1.10217 \times 10^{-2} $. Reynolds number ${\rm Re} =
1157$. Figure from Ref.\ \cite{AK2002}, computed by S. Ansumali.
}}
\end{figure}

In another numerical experiment, the ELBGK method was  tested in
the setup of the two-dimensional \index{Poiseuille flow}Poiseuille
flow. The time evolution of the computed profile as compared to
the analytical result obtained from the incompressible
Navier--Stokes equations is demonstrated in Fig.\ \ref{LBFig3}.

\begin{figure}
\includegraphics[scale=1.9]{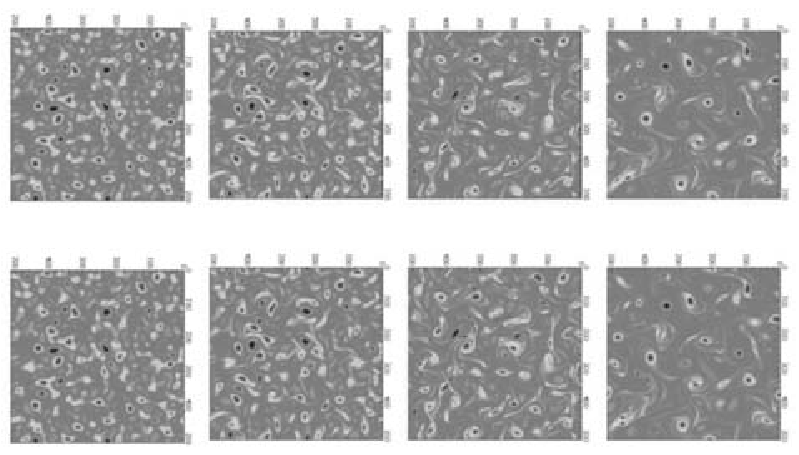} \caption{\label{turbo}
Snapshots of the vorticity field in the freely decaying $2D$
turbulence at $t=0$, $t=1000$, $t=5000$, $t=20000$ (from left to
right). Time measured in the lattice units. Eddy turnover time
$t_{\rm eddy}\approx 700$. The Reynolds number based on the mean
initial kinetic energy $E$ and the box-length $L$ equals ${\rm Re}
= L \sqrt{2E}/\nu=13134$. Upper row: Spectral method; Bottom row:
Entropic lattice Boltzmann method. Both computations were
performed on the grid of the same size ($512\times 512$ grid
points). Figure courtesy S.\ Ansumali. }
\end{figure}


The entropic lattice Boltzmann method upgrades the standard
lattice BGK scheme \cite{SucciBook01} to efficient, accurate and
unconditionally stable simulation algorithm  for high Reynolds
number flows \cite{AK2003,Athesis}. As an illustration, we present
the result of comparison of the entropic lattice Boltzmann scheme
versus the accurate spectral element code in the setup of the
freely decaying two-dimensional turbulence, see Fig. \ref{turbo}.

The essential difference between the lattice Boltzmann and the
much earlier main body research on discrete velocity models
pioneered by the seminal work of Broadwell
 \cite{Broadwell} is in two points:
\begin{itemize}
\item
  In the
lattice Boltzmann, the effort is done on fixing as much as
possible of the true (known from continuum theory) Maxwellian
dependence of relevant higher-order moments on the hydrodynamic
moments with as minimum of discrete velocities as possible, and
\item The space-time discretization to allow for large time steps
(of the order of the mean free flight rather than of the
collision). This is at variance with most of the numerical
implementation of discrete velocity models using finite difference
ideology.
\end{itemize}

\subsection*{Outlook: Lattice Boltzmann and microflows}

Gas flows at the micrometer scale constitute a major portion of
contemporary fluid dynamics of engineering interest. Because of
its relevance to the engineering of micro electro-mechanical
systems (MEMS), the branch of computational fluid dynamics focused
on micro scale phenomena is often called ``microfluidics"
\cite{Karniadakis01,REVMEMS}. Microflows are characterized by the
Knudsen number, ${\rm Kn}$, which is defined as the ratio of the
mean free path of molecules $\lambda$ and the characteristic scale
$L$ of variation of hydrodynamic fields (density, momentum, and
energy). For typical flows in microdevices, ${\rm Kn}\sim
\lambda/L$ varies from ${\rm Kn}\ll 1$ (almost-continuum flows) to
${\rm Kn} \sim 1$ (weakly rarefied flows).  Another characteristic
property of microflows is that they are highly subsonic, that is,
the characteristic flow speed is much smaller than the speed of
sound. This feature is characterized by the Mach number, ${\rm
Ma}\sim u/c_{\rm s}$, where $u$ is the characteristic flow speed,
and $c_{\rm s}$ is the (isentropic) speed of sound. Thus, for
microflows, ${\rm Ma}\ll 1$.  To be more specific, typical flow
velocities are about $0.2$ m/s, corresponding to ${\rm Ma} \sim
10^{-4}$, while values of the Knudsen number range between
$10^{-4}\le Kn \le 10^{-1}$. Finally, in the majority of
applications, microflows are quasi-two-dimensional.

Theoretical studies of gas flows at finite Knudsen number have
begun several decades ago in the realm of the Boltzmann kinetic
equation. To that end, we mention pioneering contributions by
Cercignani \cite{Cercignani75}, Sone \cite{Sone}. These studies
focused on obtaining either exact solutions of the stationary
Boltzmann kinetic equation, or other model kinetic equations in
relatively simple geometries (most often, infinite or
semi-infinite rectangular ducts), or asymptotic expansions of
these solutions.

While analytical solutions are important for a qualitative
understanding of microflows, and also for the validation of
numerical schemes, they certainly do not cover all the needs of
computational fluid dynamics of practical interest. At present,
two CFD strategies for microflows are well established.

\begin{itemize}
\item  {\it Equations of continuous fluid mechanics with slip
boundary conditions.} The simplest semi-phenomenological
observation about microflows is the break down of the  no-slip
boundary condition of fluid mechanics with increasing Knudsen
number. Since  microflows are highly subsonic, this leads to the
simplest family of models, equations of incompressible or
compressible fluid dynamics supplemented by slip velocity boundary
conditions (a review can be found in \cite{Karniadakis01}). This
approach, although widely used at the early days of microfluidics,
remains phenomenological. Moreover, it fails to predict phenomena
such as non-trivial pressure and temperature profiles observed by
more microscopic approaches.

\item {\it Direct simulation of the Boltzmann kinetic equation.}
On the other extreme, it is possible to resort to a fully
microscopic picture of collisions, and to use a molecular dynamics
approach or a simplified version thereof - the Direct Simulation
Monte Carlo method of Bird (DSMC) \cite{Bird}. DSMC is sometimes
heralded as the method of choice for simulation of the Boltzmann
equation, and it has indeed proven to be robust in supersonic,
highly compressible flows with strong shock waves. However, the
highly subsonic flows at small to moderate Knudsen number is not a
``natural" domain for the DSMC simulations where it becomes
computationally intensive \cite{DSMC}.

\end{itemize}

Since semi-phenomenological computations are not reliable, and the
fully microscopic treatment is not feasible, the approach to CFD
of microflows  must rely on reduced models of the Boltzmann
equation. Two classical routes of reducing the kinetic equations
are well known, the Chapman-Enskog method and Grad's moment
method. The Chapman-Enskog method extends the hydrodynamic
description (compressible Navier-Stokes equations) to finite ${\rm
Kn}$ in the form of a Taylor series, leading to hydrodynamic
equations of increasingly higher order in the spatial derivatives
(Burnett's hydrodynamics). Grad's method extends the hydrodynamic
equations to a closed set of equations including higher-order
moments (fluxes) as independent variables. Both methods are well
suited for theoretical studies of microflows. In particular, as
was already noted by Grad, moment equations are especially well
suited for low Mach number flows.

However, applications of Grad's moment equations or of Burnett's
hydrodynamics (or of existing extensions and generalizations
thereof) to CFD of microflows are limited at present because of
several reasons. The most severe difficulty is in formulating the
boundary conditions at the reduced level. Although some studies of
boundary conditions for moment systems were initiated recently
\cite{GrmelaKarZmi02}, this problem is far from solved. The
crucial importance of the boundary condition for microflows is
actually expected. Indeed, as the rarefaction is increasing with
${\rm Kn}$, the contribution of the bulk collisions becomes less
significant as compared to the collisions with the boundaries, and
thus the realistic modelling of the boundary conditions becomes
increasingly important.

The entropic lattice Boltzmann method seems to be a  promising
approach to simulations of microflows, and is currently an active
area of research \cite{NIE,LIM,SLIP,KWOK}. In contrast to Grad's
method, ELBM is much more compliant with the boundary conditions
(see above the diffusion wall approximation, which was also
rediscovered in \cite{Niu04}, where ELBM simulations were tested
against molecular dynamic simulations with a good agreement).
Interested reader is directed to two recent papers
\cite{AK2004a,AK2004b} where relations between the Grad and the
lattice Boltzmann constructions are considered in more detail.



\section{Concluding remarks}\label{conclusion}

The aim of this review was to give a bird-eye picture of the
method of moments pioneered by Harold Grad a half century ago.
Three relatively new  issues pertinent to the question what
physics is beyond Grad's moment approximation and how to obtain
this physics were discussed in some detail: ``Other variables''
(triangle entropy method), invariance corrections, and lifting
Grad's equations to a kinetic model. We believe that further
development of Grad's approach along the lines indicated here will
be beneficial to emerging fields of fluid dynamics, and this
review ``will be of value for both engineers and mathematicians
... who may attempt to turn the invariance condition equation into
rigorous mathematics", as was suggested by the referee of this
paper. As per mathematical rigor, the situation is at least not
hopeless for finite-dimensional systems, such as ordinary
differential equations of chemical kinetics. However, much more
work is needed for infinite-dimensional systems like the Boltzmann
equation where the present level of mathematical achievements in
such things as existence and uniqueness of solutions does not
allow even to start the rigorous talking about construction of
invariant manifolds. Some mathematical requirements are formulated
in the recent book \cite{GKbook2}.

\section*{Acknowledgement}

It is our pleasure to thank Dr. S. Ansumali for his contributions
to section \ref{LBsection}, and for providing the figures. IVK was
supported by the Swiss Federal Department of Energy (BFE) under
the project Nr. 100862 ``Lattice Boltzmann simulations for
chemically reactive systems in a micrometer domain".


\begin{thebibliography}{99}

\bibitem{Perthame2002}
P.~Andries, K.~Aoki, and B.~Perthame.
 A consistent {BGK}-type model for gas mixtures.
 {\em J. Stat. Phys.}, 106:993--1018, 2002.

\bibitem{kirp}
S.~Ansumali, S.~S. Chikatamarla, C.~M. Frouzakis, and K.~Boulouchos.
 {E}ntropic lattice {B}oltzmann simulation of the flow past square
  cylinder.
 {\em Int. J. Mod. Phys. C}, 15(3):435--445, 2004.

\bibitem{AK2004b}
S.~Ansumali, Ch.~E. Frouzakis, I.~V. Karlin, and I.~G. Kevrekidis.
 A glimplse into hydrodynamics at the microscale.
 {\em preprint}, 2004.

\bibitem{AK2002}
S~Ansumali and I.~V. Karlin.
 Entropy function approach to the lattice {B}oltzmann method.
 {\em J. Stat. Phys.}, 107:291--308, 2002.

\bibitem{AK2002b}
S.~Ansumali and I.~V. Karlin.
 {K}inetic boundary condition for the lattice {B}oltzmann method.
 {\em Phys. Rev. E}, 66:026311(1--6), 2002.

\bibitem{AK2002a}
S~Ansumali and I.~V. Karlin.
 Single relaxation time model for entropic lattice {B}oltzmann
  methods.
 {\em Phys. Rev. E}, 65:056312(1--9), 2002.

\bibitem{AK2004a}
S.~Ansumali, I.~V. Karlin, Ch.~E. Frouzakis, and K.~B. Boulouchos.
 Entropic lattice {B}oltzmann method for microflows.
 {\em http://xxx.lanl.gov/abs/cond-mat/0412555}, 2004.

\bibitem{AKOe2003}
S.~Ansumali, I.~V. Karlin, and H.~C. \"{O}ttinger.
 Minimal entropic kinetic models for simulating hydrodynamics.
 {\em Europhys. Lett.}, 63:798--804, 2003.

\bibitem{Athesis}
S.~Ansumali.
 {\em Minimal kinetic modeling of hydrodynamics}.
 PhD thesis, Swiss Federal Institute of Technology Z\"{u}rich, 15534,
  2004.

\bibitem{Bird}
G.~A. Bird.
 {\em Molecular {G}as {D}ynamics and the {D}irect {S}imulation of
  {G}as {F}lows. {T}heory and {A}pplication of the {B}oltzmann {E}quation}.
 Clarendon Press, Oxford, 1994.

\bibitem{Karniadakis01}
A.~Beskok and G.~E. Karniadakis.
 {\em Microflows: {F}undamentals and {S}imulation}.
 Springer, Berlin, 2001.

\bibitem{Bobylev82}
A.~V. Bobylev.
 On the {C}hapman-{E}nskog and {G}rad methods.
 {\em Dokl. Acad. Nauk SSSR}, 262:71, 1982.

\bibitem{Broadwell}
J.~E. Broadwell.
 Study of rarefied shear flow by the discrte velocity method.
 {\em J. Fluid Mech.}, 19:401--414, 1964.

\bibitem{Boghosian01}
B.~M. Boghosian, J~Yepez, P.~V. Coveney, and A.~J. Wagner.
 {E}ntropic lattice {B}oltzmann methods.
 {\em Proc. Roy. Soc. Lond. A}, 457:717--766, 2001.

\bibitem{Chapman}
S.~Chapman and T.~G. Cowling.
 {\em The {M}athematical {T}heory of {N}on-{U}niform {G}ases}.
 Cambridge University Press, Cambridge, 1970.

\bibitem{Cercignani75}
C.~Cercignani.
 {\em Theory and {A}pplication of the {B}oltzmann {E}quation}.
 Scottish Academic Press, Edinburgh, 1975.

\bibitem{GGK04}
A.~N. Gorban, P.~A. Gorban, and I.~V. Karlin.
 Legendre integrators, postprocessing and quasiequilibrium.
 {\em J. Non-Newtonian Fluid Mech.}, 120:149--167, 2004.

\bibitem{GK91}
A.~N. Gorban and I.~V. Karlin.
 Quasi-equilibrium approximations and non-standard expansions in the
  theory of the {B}oltzmann kinetic equation.
 In R.~G. Khlebopros, editor, {\em Mathematical Modeling in Biology
  and Chemistry (New Approaches)}, pages 69--117. Nauka, Novosibirsk, 1991.
 English translation of the first part of this paper (triangle entropy
  method): http://arXiv.org/abs/cond-mat/0305599.

\bibitem{GK91a}
A.~N. Gorban and I.~V. Karlin.
 Structure and approximations of the {C}hapman-{E}nskog expansion for
  {G}rad linearized equations.
 {\em Sov. Phys. JETP}, 73(4):637--641, 1991.

\bibitem{GK92}
A.~N. Gorban and I.~V. Karlin.
 Thermodynamic parameterization.
 {\em Physica A}, 190:393--404, 1992.

\bibitem{GKbook1}
A.~N. Gorban and I.~V. Karlin.
 {\em {N}ew {M}ethods for {S}olving the {B}oltzmann {E}quations},
  volume 10 [Physical Kinetics] of {\em Scientific Siberian A}.
 AMSE Press, Tassin, 1993.

\bibitem{GK94bgk}
A.~N. Gorban and I.~V. Karlin.
 General approach to constructing models of the {B}oltzmann equation.
 {\em Physica A}, 206:401--420, 1994.

\bibitem{GK94inv}
A.~N. Gorban and I.~V. Karlin.
 Method of invariant manifolds and regularization of acoustic spectra.
 {\em Transport Theory and Stat. Phys.}, 23:559--632, 1994.

\bibitem{GK96}
A.~N. Gorban and I.~V. Karlin.
 Scattering rates versus moments: {A}lternative {G}rad equations.
 {\em Phys. Rev. E}, 54:R3109--R3112, 1996.

\bibitem{GK96a}
A.~N. Gorban and I.~V. Karlin.
 Short-wave limit of hydrodynamics: {A} soluble example.
 {\em Phys. Rev. Lett.}, 77:282--285, 1996.

\bibitem{GK2002}
A.~N. Gorban and I.~V. Karlin.
 Method of invariant manifold for chemical kinetics.
 {\em Chem. Eng. Sci.}, 58:4751--4768, 2003.

\bibitem{GKbook2}
A.~N. Gorban and I.~V. Karlin.
 {\em Invariant {M}anifolds for {P}hysical and {C}hemical {K}inetics},
  volume 660 of {\em Lect. Notes Phys.}
 Springer, Berlin, 2004.

\bibitem{GK04b}
A.~N. Gorban and I.~V. Karlin.
 Uniqueness of thermodynamic projector and kinetic basis of molecular
  individualism.
 {\em Physica A}, 336(3-4):391--432, 2004.
 Preprint online: http://arxiv.org/abs/cond-mat/0309638.

\bibitem{GrmelaKarZmi02}
M.~Grmela, I.~V. Karlin, and V.~B. Zmievski.
 Boundary layer variational principle: {A} case study.
 {\em Phys. Rev. E}, 66:011201(1--12), 2002.

\bibitem{GKZ2004}
A.~N. Gorban, I.~V. Karlin, and A.~Yu. Zinovyev.
 {C}onstructive methods of invariant manifolds for kinetic problems.
 {\em Physics Reports}, 396:197--403, 2004.

\bibitem{GKZ2003}
A.~N. Gorban, I.~V. Karlin, and A.~Yu. Zinovyev.
 Invariant grids for reaction kinetics.
 {\em Physica A}, 333:106--154, 2004.

\bibitem{GKZD2000}
A.~N. Gorban, I.~V. Karlin, V.~B. Zmievskii, and S.~V. Dymova.
 Reduced description in reaction kinetics.
 {\em Physica A}, 275:361--379, 2000.

\bibitem{obkhod}
A.~N. Gorban.
 {\em Equilibrium Encircling. Equations of Chemical Kinetics and their
  Thermodynamic Analysis}.
 Nauka, Novosibirsk, 1984.

\bibitem{Grad49}
H.~Grad.
 On the kinetic theory of rarefied gases.
 {\em Comm.\ Pure Appl.\ Math.}, 2:331--407, 1949.

\bibitem{REVMEMS}
C.-M. Ho and Y.-C. Tai.
 Micro-electro-mechanical-systems({MEMS}) and fluid flows.
 {\em Annu. Rev. Fluid Mech.}, 30:579--612, 1998.

\bibitem{Ilg2003}
P.~Ilg, I.~V. Karlin, M.~Kr\"{o}ger, and H.~C. \"{O}ttinger.
 Canonical distribution functions in polymer dynamics: {II}.
  {L}iquid-crystalline polymers.
 {\em Physica A}, 319:134--150, 2003.

\bibitem{Ilg2002}
P.~Ilg, I.~V. Karlin, and H.~C. \"{O}ttinger.
 Canonical distribution functions in polymer dynamics: I. {D}ilute
  solutions of flexible polymers.
 {\em Physica A}, 315(3-4):318--336, 2002.

\bibitem{AK2003}
I.~V. Karlin, S.~Ansumali, E.~De~Angelis, H.~C. \"{O}ttinger, and
S.~Succi.
 Entropic lattice {B}oltzmann method for large scale turbulence
  simulation.
 {\em http://xxx.lanl.gov/abs/cond-mat/0306003}, 2003.

\bibitem{Karlin86}
I.~V. Karlin.
 Relaxation of chemical reaction rates under translationally
  nonequilibrium conditions.
 In {\em Proc. VIII USSR Symp. on Burning and Combustion}, pages
  97--99, Chernogolovka, 1986. Inst. Chem. Phys.

\bibitem{Kdiss}
I.~V. Karlin.
 {\em Method of Invariant Manifold in Kinetic Theory}.
 PhD thesis, AMSE University, Tassin, 1992.

\bibitem{KDN97}
I.~V. Karlin, G.~Dukek, and T.~Nonnenmacher.
 Invariance principle for extension of hydrodynamics: Nonlinear
  viscosity.
 {\em Phys. Rev. E}, 55(2):1573--1576, 1997.

\bibitem{KFOe99}
I.~V. Karlin, A~Ferrante, and H.~C. \"{O}ttinger.
 Perfect entropy functions of the lattice {B}oltzmann method.
 {\em Europhys. Lett.}, 47:182--188, 1999.

\bibitem{Karlin2002a}
I.~V. Karlin and A.~N. Gorban.
 Hydrodynamics from {G}rad's equations: {W}hat can we learn from exact
  solutions?
 {\em Ann. Phys. (Leipzig)}, 11(10-11):783--833, 2002.

\bibitem{Karlin98}
I.~V. Karlin, A.~N. Gorban, G.~Dukek, and T.~Nonnenmacher.
 Dynamic correction to moment approximations.
 {\em Phys. Rev. E}, 57:1668--1672, 1998.

\bibitem{KGSB98}
I.~V. Karlin, A.~Gorban, S.~Succi, and V.~Boffi.
 Maximum entropy principle for lattice kinetic equations.
 {\em Phys. Rev. Lett.}, 81:6--9, 1998.

\bibitem{Kogan65}
A.~M. Kogan.
 {D}erivation of {G}rad-type equations and study of their properties
  by the method of entropy maximization.
 {\em Prikl. Math. Mech.}, 29:122--133, 1965.

\bibitem{Levermore96}
C.~D. Levermore.
 Moment closure hierarchies.
 {\em J. Stat. Phys.}, 83:1021, 1996.

\bibitem{Lewis67}
R.~M. Lewis.
 A unified principle in statistical mechanics.
 {\em J. Math. Phys.}, 8:1448--1459, 1967.

\bibitem{KWOK}
B.~Li and D.~Kwok.
 Discrete {B}oltzmann equation for microfluidics.
 {\em Phys. Rev. Lett.}, 90:124502, 2003.

\bibitem{LIM}
C.~Y. Lim, C.~Shu, X.~D. Niu, and Y.~T. Chew.
 Application of lattice {B}oltzmann method to simulate microchannel
  flows.
 {\em Phys. Fluid}, 107:2299--2308, 2002.

\bibitem{Muller}
I.~M\"{u}ller and T.~Ruggeri.
 {\em Extended {T}hermodynamics}.
 Springer, Berlin, 1993.

\bibitem{NIE}
X.~Nie, G.~Doolen, and S.~Chen.
 Lattice-{B}oltzmann simulations of fluid flows in {MEMS}.
 {\em J. Stat. Phys.}, 107:279--289, 2002.

\bibitem{Niu04}
X.~D. Niu, C.~Shu, and Y.T. Chew.
 Lattice {B}oltzmann {BGK} model for simulation of micro flows.
 {\em Euro. Phys. Lett.}, 67:600--606, 2004.

\bibitem{DSMC}
E.~S. Oran, C.~K. Oh, and B.~Z. Cybyk.
 {D}irect simulation {M}onte {C}arlo: {R}ecent advances and
  applications.
 {\em Annu. Rev. Fluid Mech.}, 30:403--441, 1998.

\bibitem{Santos2000}
A.~Santos.
 Nonlinear viscosity and velocity distribution function in a simple
  longitudinal flow.
 {\em Phys. Rev. E}, 62:4747--4751, 2000.

\bibitem{He98}
X.~Shan and X.~He.
 Discretization of the velocity space in the solution of the
  {B}oltzmann equation.
 {\em Phys. Rev. Lett.}, 80:65--68, 1998.

\bibitem{Slemrod98}
M.~Slemrod.
 Renormalization of the {C}hapman-{E}nskog expansion: {I}sothermal
  fluid flow and {R}osenau saturation.
 {\em J. Stat. Phys.}, 91:285--305, 1998.

\bibitem{Sone}
Y.~Sone.
 {\em Kinetic {T}heory and {F}luid {D}ynamics}.
 Birkh\"{a}user, Basel, 2002.

\bibitem{Struchtrup2002a}
H.~Struchtrup and M.~Torrilhon.
 Regularization of {G}rad's 13 moment equations: {D}erivation and
  linear analysis.
 {\em Phys. Fluids}, 15:2668--2680, 2003.

\bibitem{SucciBook01}
S.~Succi.
 {\em The {L}attice {B}oltzmann {E}quation for {F}luid {D}ynamics and
  {B}eyond}.
 Oxford University Press, Oxford, 2001.

\bibitem{SLIP}
S.~Succi.
 Mesoscopic modeling of slip motion at fluid-solid interfaces with
  heterogeneous catalysis.
 {\em Phys. Rev. Lett.}, 89:064502, 2002.

\bibitem{ZKD00}
V.~B. Zmievskii, I.~V. Karlin, and M.~Deville.
 The universal limit in dynamics of dilute polymeric solutions.
 {\em Physica A}, 275(1-2):152--177, 2000.

\end{thebibliography}
\end{document}